\documentclass[aps,prd,onecolumn,superscriptaddress,nofootinbib]{revtex4-2}
\usepackage{blindtext}
\usepackage{titlesec}
\usepackage{amsfonts}
\usepackage{amsmath}
\usepackage{amssymb}
\usepackage{multirow}
\usepackage{array} \newsavebox\cellbox
\usepackage{calc}
\usepackage{float}
\usepackage{mathtools}
\usepackage[usenames, dvipsnames, table]{xcolor}
\usepackage{booktabs}
\usepackage{makecell, cellspace}
\usepackage{color, colortbl}
\usepackage[caption=false]{subfig}
\usepackage{hyperref}
\hypersetup{
    colorlinks,
    linkcolor={red!60!black},
    citecolor={green!50!black},
    urlcolor={blue!70!black}
}
\usepackage{tikz}
\usepackage{pgfplots}
\usetikzlibrary{fadings, calc}
\tikzset{
  treenode/.style = {shape=rectangle, rounded corners,
                     draw, align=center,
                     top color=white, bottom color=blue!20},
  root/.style     = {treenode, font=\small, bottom color=red!30},
  tbd/.style      = {treenode, bottom color=gray, font=\tiny},
  env/.style      = {treenode, font=\tiny},
  dummy/.style    = {circle,draw}
}

\def\app#1#2{%
  \mathrel{%
    \setbox0=\hbox{$#1\sim$}%
    \setbox2=\hbox{%
      \rlap{\hbox{$#1\propto$}}%
      \lower1.1\ht0\box0%
    }%
    \raise0.25\ht2\box2%
  }%
}
\def\approxprop{\mathpalette\app\relax}

\newcolumntype{q}[2]{%
>{\begin{lrbox}\cellbox}%
    l%
    <{\end{lrbox}%
\makebox[#2][#1]{\usebox\cellbox}}}

\newlength\mylen
\newcommand\coltwo[2]{%
   \multicolumn{2}{q{#1}{\dimexpr2\mylen+2\arraycolsep}}{#2}}
\newcommand\colthree[2]{%
   \multicolumn{3}{q{#1}{\dimexpr3\mylen+4\arraycolsep}}{#2}}
   
\pgfplotsset{compat=1.15}

\begin{document}

\title{Solving nonlinear Klein-Gordon equations on unbounded domains via the Finite Element Method}
\author{Hugo L\'{e}vy}
\email{hugo.levy@onera.fr}
\affiliation{DPHY, ONERA, Universit\'{e} Paris Saclay
F-92322 Ch\^{a}tillon - France}
\affiliation{Sorbonne Universit\'{e}, CNRS, UMR 7095, Institut d’Astrophysique de Paris, 98 bis bd Arago, 75014 Paris, France}
\author{Jo\"{e}l Berg\'{e}}
\affiliation{DPHY, ONERA, Universit\'{e} Paris Saclay
F-92322 Ch\^{a}tillon - France}
\author{Jean-Philippe Uzan}
\affiliation{Sorbonne Universit\'{e}, CNRS, UMR 7095, Institut d’Astrophysique de Paris, 98 bis bd Arago, 75014 Paris, France}

\begin{abstract}
A large class of scalar-tensor theories of gravity exhibit a screening mechanism that dynamically suppresses fifth forces in the Solar system and local laboratory experiments. Technically, at the scalar field equation level, this usually translates into nonlinearities which strongly limit the scope of analytical approaches.
This article presents \emph{femtoscope} \textemdash \ a Python numerical tool based on the Finite Element Method (FEM) and Newton method for solving Klein-Gordon-like equations that arise in particular in the symmetron or chameleon models. Regarding the latter, the scalar field behavior is generally only known infinitely far away from the its sources. We thus investigate existing and new FEM-based techniques for dealing with asymptotic boundary conditions on finite-memory computers, whose convergence are assessed. Finally, \emph{femtoscope} is showcased with a study of the chameleon fifth force in Earth orbit.
\end{abstract}

\maketitle


\section{Introduction}
\label{sec:intro}

General Relativity (GR) is our best understanding of gravity. It passes all the tests thrown at it so far \cite{gr-tests}, including the most recent test of the weak equivalence principle \textemdash \ the MICROSCOPE experiment \cite{results-PRL, results-CQG}. However, gravity cannot go but hand in hand with cosmology, a discipline that relates to many questions in fundamental physics. In the standard model of cosmology, most of our Universe's mass-energy budget is made up of an understood dark sector, namely dark matter and dark energy, which may be seen as necessary patches to account for cosmological observations while assuming the validity of GR \cite{Jain:2013wgs}. Precision tests of gravity on astrophysical and cosmological scales (either from the large scale structures~\cite{Uzan:2000mz,Uzan:2010ri} or the test of the equivalence principle with fundamental constants~\cite{Uzan:2002vq,Uzan:2010pm}) have been developed but have not provided precise enough insight, opening the way to a large activity on the so-called modified gravity theories, i.e. gravity theories beyond GR. One of the simplest extensions of GR yet phenomenologically rich consists in supplementing the metric field with additional fields. Among these, scalar fields have been vastly studied as they could provide possible models in cosmology, from inflation \cite{inflation_scalar} to late time cosmic acceleration \cite{ratra-peebles-1988, caldwell-1998, quintessence-review, chameleon-dark-energy}. Untowardly, such scalar-tensor models are hardly viable because they mediate a so-called ``$\mathrm{5^{th}}$-force" that has not been detected so far (see Ref.~\cite{Uzan:2020aig} for recent ideas to detect them in the lab). As a consequence, most of these models are already severely constrained by the various experimental tests of gravity below the Solar system scale unless they are attracted toward GR during the cosmological evolution~\cite{Damour:1994zq}.

Some models remain viable via \textit{screening mechanisms}. Screening mechanisms suppress the $\mathrm{5^{th}}$-force in Earth and Solar system based experiments, letting them constrained but still viable while offering a fruitful phenomenology for the cosmic acceleration on astrophysical scales. The chameleon field is an example of a dynamically screened scalar field, for which screening arises from the local density dependence of the field's mass \cite{khoury_weltman_original, surprises}. The chameleon mechanism has been extensively tested, see Refs.~\cite{Burrage2018, lab-tests} for comprehensive reviews. As a result, entire regions of the chameleon parameter space are already ruled out, mainly thanks to laboratory experiments \cite{constraints}. Exploring unconstrained parts of the parameter space thus requires designing innovative experiments, which is partly impeded by the difficulty to accurately model the field behavior.

At the equations level, the screening mechanism relies on nonlinearities in the partial differential equation (PDE) governing the chameleon field's dynamics \textemdash \ the Klein-Gordon equation. Consequently, analytical approaches are of little help to derive quantitative information about the chameleonic fifth force, although some approximations in the case of highly symmetrical setups are worth mentioning \textemdash \ e.g. Ref.~\cite{khoury_weltman_original} for homogeneous solid sphere immersed in lower density background, Ref.~\cite{chameleon-ellipse} for ellipsoidal sources, or Ref.~\cite{martin_2020_PRD} for non-coaxial nested cylinders (semi-analytical). The nonlinear nature of the PDE as well as the need to study the chameleon field profile around more diverse matter distributions tilt the balance in favor of numerical simulations. In particular, the finite element method (FEM) is well-suited for that purpose as it hinges on meshes that can fit virtually any given geometry. The recent Python code SELCIE performs this ambitious task \cite{Briddon_2021}. In Ref.~\cite{martin_2019_PRD}, the finite difference technique is employed in 1D to study the chameleon mechanism in the context of the MICROSCOPE experiment, while Ref.~\cite{fdm_cham} conducts similar simulations in atom interferometry setups. Other types of screening have been numerically investigated, see e.g. Refs.~\cite{Braden_2021, fem_symmetron}.

PDEs cannot be mathematically well-posed unless specified with proper boundary (and initial) conditions. This is a burning issue in the case of the chameleon field as the theory does not predict its exact behavior anywhere near the matter sources in the general case. Instead, it is known only infinitely far away from the sources, where density no longer fluctuates so that the scalar field relaxes to the value that minimizes some effective potential in this remote region. This issue is often overlooked, or at least circumvented by setting boundary conditions at finite distance \cite{Briddon_2021} which is not legitimate at all in the general case. Indeed, the assumption used to justify the choice of setting a boundary condition at a finite distance is either that: (i) the setup is encapsulated into walls thick enough that they are screened, i.e. the unknown field takes the value that minimizes the effective potential deep inside the walls (e.g. Refs.~\cite{martin_2020_PRD, fdm_cham, shape_dependence}); or (ii) the boundary of the numerical domain is set sufficiently far away from the matter sources (typically several Compton wavelengths away) so that the field has almost reached its asymptotic value \cite{nfw_halos}.
None of these two hypotheses are fully satisfactory since they cannot be true in all regions of the parameter space. This difficulty has already been pointed out in Ref.~\cite{martin_2019_PRD} which managed to circumvent it through the use of a shooting method. However, this approach was limited to 1D cases symmetrical about the origin.

This work aims at overcoming the challenge of accurately representing the asymptotic behavior of the chameleon field at infinity while dealing with arbitrary matter distribution setups. This has not been achieved in any of the aforementioned codes, which is the driving motivation to develop the new code \emph{femtoscope}\footnote{This name was chosen to 1) echo the MICROSCOPE space mission, 2) contain \textit{FEM} which is the commonly adopted acronym for `Finite Element Method' and 3) contain the Danish prefix \textit{femto-} $\to 10^{-15}$.}. This software package builds on top of \textit{Sfepy} \cite{sfepy}, a FEM Python package. The Klein-Gordon equation governing the field dynamics being nonlinear, \textit{semilinear} to be specific, a Newton solver is employed with the possibility to activate a line search algorithm at each iteration for enhanced convergence. The bulk of the work lies in the implementation of asymptotic boundary conditions on the unknown scalar field. To that extent, three distinct techniques are introduced: one relies on a compactification of the original unbounded domain while the two others are based on a domain splitting followed by a Kelvin inversion. Such techniques are very general and may interest other fields of study. This article showcases their use on PDEs relevant to us, namely Poisson equation for deriving the gravitational potential and Klein-Gordon equation for studying the chameleon field. The former has closed-form solutions in some simple cases which will serve as a basis for validating the implementation.

This article is organized as follows. First, Sec.~\ref{sec:model} describes the chameleon model, from its physics to the mathematical problem to be tackled. Sec.~\ref{sec:methods} is dedicated to a thorough description of the numerical tools involved in \emph{femtoscope}, with emphasis laid on the handling of asymptotic boundary conditions together with a technical review. After a brief overview of \emph{femtoscope}'s workflow in Sec.~\ref{sec:femtoscope}, we present the results of a first study of the chameleon $\mathrm{5^{th}}$-force effects in terrestrial orbit using realistic density models in Sec.~\ref{sec:chameleon-earth}. Lastly, Sec.~\ref{sec:conclusion} concludes and draws future studies with \emph{femtoscope}.

\section{From the chameleon model to the asymptotic boundary value problem}
\label{sec:model}

This section recalls the basis for the chameleon model without elaborating too much on the physical side (see Refs.~\cite{khoury_weltman_original, Burrage2018} for that purpose). Ultimately, our goal is to write down a mathematically well-posed problem, which will pave the way to the numerical techniques implemented in this study. For clarity, we work in natural units where the speed of light $c$ and the reduced Planck constant $\hbar$ are set to unity.

\subsection{The chameleon field}
\label{subsec:cham-physics}

In the Einstein frame, the action of a generic scalar-tensor theory is
\begin{equation}
\small
    S_{\mathrm{cham}} \equiv \int \mathrm{d}^4 x \sqrt{-g} \left[ \frac{\mathrm{M_{Pl}^2}}{2} R - \frac{1}{2} g^{\mu \nu} \partial_{\mu} \phi \partial_{\nu} \phi - V(\phi) \right] + \int \mathrm{d}^4 x \mathcal{L}_{\mathrm{m}} \left( \Omega^2_{(i)}(\phi) g_{\mu \nu}, \psi_{\mathrm{m}}^{(i)} \right) \, ,
\end{equation}
where $\mathrm{M_{Pl}} \equiv 1/\sqrt{8 \pi G}$ is the reduced Planck mass, $V$ is the bare potential of the scalar field $\phi$ (which completely dictates its dynamics if it were not coupled to matter), $\mathcal{L}_{\mathrm{m}}$ is the matter Lagrangian for the matter fields $\psi_{\mathrm{m}}^{(i)}$ which couple to $\phi$ through the conformal factors $\Omega_{(i)}(\phi)$ respectively. $R$, $g_{\mu \nu}$, $g$ are respectively the Einstein frame's Ricci scalar, the metric tensor and its determinant, assuming signature $(-,+,+,+)$. While the field could have different couplings to each matter component, we restrict our analysis to a universal coupling allowing us to drop the index $i$. It follows that we can introduce a unique Jordan frame metric, $\Tilde g_{\mu\nu}=\Omega^2 g_{\mu\nu}$. The chameleon model is then completely defined by the two functions of $\phi$, $\Omega(\phi)$ and  $V(\phi)$. In the following, we choose the specific form of the former to be
\begin{equation}
\Omega(\phi) = \mathrm{e}^{\frac{\beta}{\mathrm{M_{Pl}}} \phi} \, ,
\end{equation}
with $\beta$ a dimensionless coupling constant. For the latter, we use the Ratra-Peebles inverse power-law potential of energy scale $\Lambda$ and exponent $n$,
\begin{equation}
    V(\phi) = \Lambda^4 \left( 1 + \frac{\Lambda^n}{\phi^n} \right) \, .
\end{equation}
The dynamics of the chameleon field is then obtained by varying the action with respect to $\phi$, yielding the Klein-Gordon equation
\begin{equation}
    \partial^{\mu} \partial_{\mu} \phi = \frac{\mathrm{d}V}{\mathrm{d}\phi} - \frac{\beta}{\mathrm{M_{Pl}}} \mathrm{e}^{\frac{4\beta}{\mathrm{M_{Pl}}} \phi} T^{\mu \nu} \Tilde{g}_{\mu \nu} \, ,
\label{eqn:kg1}
\end{equation}
where $T_{\mu \nu}$ is the stress-energy tensor defined as
$$ T_{\mu \nu} = - \frac{2}{\sqrt{-g}} \frac{\delta (\sqrt{-g} \mathcal{L}_{\mathrm{m}})}{\delta g^{\mu \nu}} \, . $$
In the Newtonian limit the field's Klein-Gordon equation reduces to
\begin{equation}
    \Box \phi = \frac{\mathrm{d}V_{\mathrm{eff}}}{\mathrm{d}\phi} = \frac{\beta}{\mathrm{M_{Pl}}} \rho \mathrm{e}^{\frac{\beta \phi}{\mathrm{M_{Pl}}}} - \frac{n \Lambda^{n+4}}{\phi^{n+1}} \, ,
\label{eqn:kg2}
\end{equation}
with $\Box$ the d'Alembert operator and
\begin{equation}
V_{\mathrm{eff}} = V(\phi) + \rho \exp(\beta \phi / \mathrm{M_{Pl}})
\label{eqn:effective_potential}
\end{equation}
the effective potential. In this article, we only consider static configurations of matter (there is a discussion of the quasi-static approximation in Ref.~\cite{numerical_modgravity} section IV and references therein), so that the d'Alembertian reduces to the Laplacian. The final form of the Klein-Gordon equation to be studied is thus
\begin{equation}
    \Delta \phi = \frac{\beta}{\mathrm{M_{Pl}}} \rho - \frac{n \Lambda^{n+4}}{\phi^{n+1}} \, ,
\label{eqn:dimKG}
\end{equation}
where we have further assumed $\beta \phi \ll \mathrm{M_{Pl}}$ to get rid of the exponential term in Eq.~(\ref{eqn:kg2}). The geodesic equation allows to identify the effect of the scalar field and hence defines the chameleon fifth force experienced by a point-mass of mass $m$ as
\begin{equation}
    \mathbf{F}_{\phi} = -m \frac{\beta}{\mathrm{M_{Pl}}} \boldsymbol{\nabla} \phi \, ,
\label{eqn:force-natural}
\end{equation}
which is the central physical quantity in this study as it could be measured at length scales below the Solar system scale. Note that Eq.~(\ref{eqn:force-natural}) is an approximation of the fifth force, whose exact expression can be found in e.g. Ref.~\cite{Uzan:2020aig}. Additionally, the reader interested in the classification of modified gravity models and the nature of their underlying equation may refer to Table~1 of Ref.~\cite{numerical_modgravity}.

\subsection{Behavior of the field far away from matter sources}
\label{subsec:asymptotic-bc-phys}

Now, we assume that the density uniformly decays to some vacuum density $\rho_{\mathrm{vac}}$ at infinity. The chameleon field then relaxes to the value $\phi_{\mathrm{vac}}$ that minimizes the effective potential $V_{\mathrm{eff}}$ (\ref{eqn:effective_potential}). Again, assuming that $\beta \phi \ll \mathrm{M_{Pl}}$ yields
\begin{equation}
    \phi_{\mathrm{vac}} \simeq \left( \mathrm{M_{Pl}} \frac{n \Lambda^{n+4}}{\beta \rho_{\mathrm{vac}}} \right)^{\frac{1}{n+1}} \, .
\label{eqn:phi_vac}
\end{equation}
It is thus reasonable to impose the asymptotic condition $$\phi(r, \theta, \varphi) \underset{r \to +\infty}{\longrightarrow} \phi_{\mathrm{vac}} \, , $$
where $(r, \theta, \varphi)$ are spherical coordinates. Indeed, by construction, this asymptotic value of the field makes the right-hand-side (r.h.s.) of equation (\ref{eqn:dimKG}) vanish at infinity.

In many articles dealing with the Klein-Gordon equation (\ref{eqn:dimKG}), an additional asymptotic condition is enforced on the field's gradient at infinity \cite{martin_2019_PRD, unnecessary_1, unnecessary_2}, namely
\begin{equation}
    \| \boldsymbol{\nabla} \phi \| \underset{\|\mathbf{x}\| \to +\infty}{\longrightarrow} 0 \, .
\label{eqn:vanishing-gradient}
\end{equation}
Yet, we can actually show that if $\phi : \mathbb{R}^3 \to \mathbb{R}$ satisfies Eq.~(\ref{eqn:dimKG}) and is such that $\partial_{\theta} \phi$, $\partial_{\varphi} \phi$, $\partial_{\theta}^2 \phi$, $\partial_{\varphi}^2 \phi$ are $O(1)$ as $r \to +\infty$, then Eq.~(\ref{eqn:vanishing-gradient}) is granted (see Appendix \ref{sec:bc-proof} for the proof of that statement). Nevertheless, this remark is of minor importance for the numerical techniques to be introduced in Sec.~\ref{sec:methods}. Indeed, the vanishing gradient condition (\ref{eqn:vanishing-gradient}) naturally arises in the framework of FEM.

\subsection{Dimensionless version of the Klein-Gordon equation}

Eq.~(\ref{eqn:dimKG}) is seemingly governed by the three parameters $(\beta, \Lambda, n)$. The nondimensionalisation of this equation is done as in Ref.~\cite{Briddon_2021}, that is we set:
\begin{itemize}
    \item[--] $\rho_0$ a characteristic density of the problem (for instance the vacuum density);
    \item[--] $\phi_0 \equiv \left( \mathrm{M_{Pl}} \frac{n \Lambda^{n+4}}{\beta \rho_{\mathrm{0}}} \right)^{\frac{1}{n+1}}$ the value of the field that minimizes the effective potential in a medium of density $\rho_0$;
    \item[--] $L_0$ a characteristic length scale of the system under study;
\end{itemize}
and introduce the dimensionless quantities $\hat{\rho} \equiv \rho / \rho_0$ and $\hat{\phi} \equiv \phi / \phi_0$ as well as the modified Laplacian operator $\hat{\Delta} \equiv L_0^2 \Delta$. The resulting dimensionless Klein-Gordon equation reads

\begin{equation}
    \alpha \hat{\Delta} \hat{\phi} = \hat{\rho} - \hat{\phi}^{-(n+1)} \quad \text{with} \quad \alpha \equiv \left( \frac{\mathrm{M_{Pl}} \Lambda}{L_0^2 \rho_0 \beta} \right) \left( \frac{n \mathrm{M_{Pl}}\Lambda^3}{\beta \rho_0} \right)^{\frac{1}{n+1}} \, .
\label{eqn:kg3}
\end{equation}

\begin{figure*}[b]
    \centering
    \includegraphics[width=\textwidth]{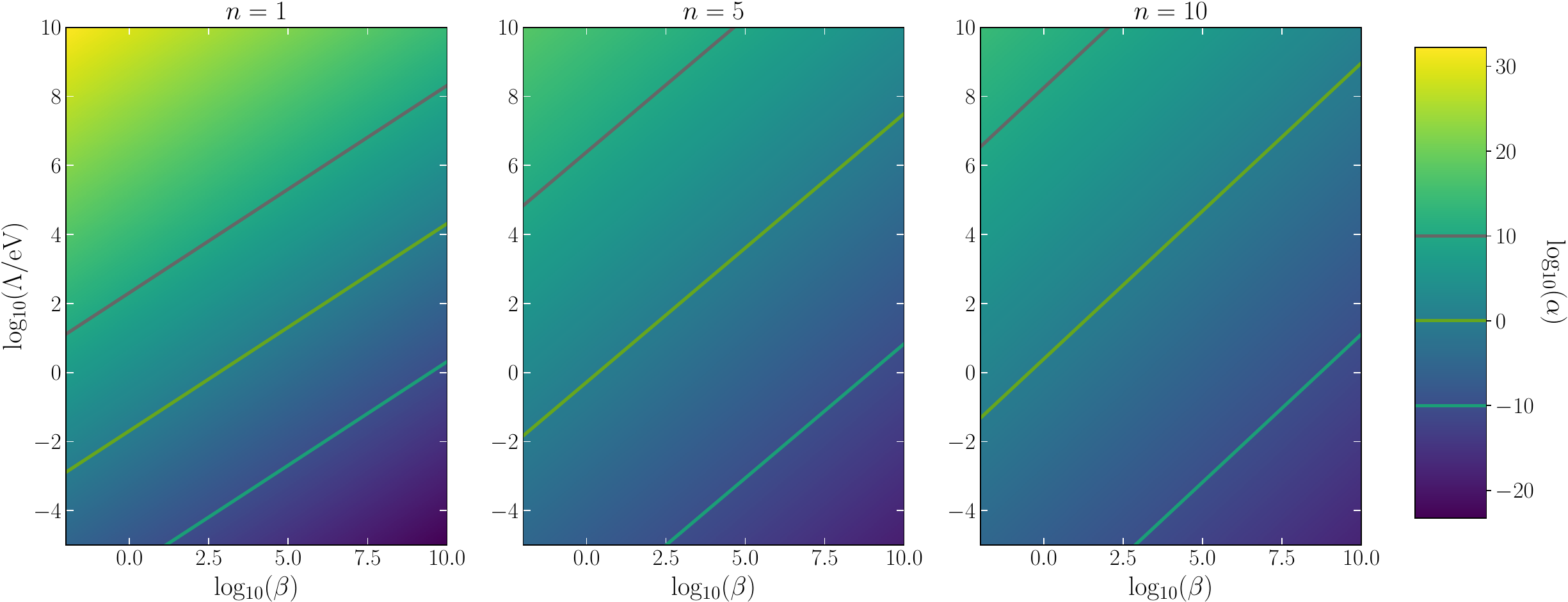}
    \caption{Mapping from the chameleon space parameter $(\beta, \Lambda)$ to the dimensionless $\alpha$ parameter appearing in the dimensionless Klein-Gordon equation (\ref{eqn:kg3}). For a given $n$, the pairs $(\beta, \Lambda)$ that map to the same $\alpha$ value result in the same chameleon field profile up to a global rescaling constant (as shown by the three ad hoc lines).}
    \label{fig:alpha}
\end{figure*}

\noindent Fig.~\ref{fig:alpha} depicts the mapping $(\beta, \Lambda) \mapsto \alpha$ for different values of the integer exponent $n$. It has to be put into perspective with Fig. 21 of Ref.~\cite{martin_2019_PRD}, which shows the delimitation between the screened and the unscreened regimes for the MICROSCOPE setup in the chameleon's parameter space. Indeed, for a given mass distribution, the chameleon dynamics is solely determined by $n$ and $\alpha$, whose iso-values are straight lines in the $(\mathrm{log}\beta, \mathrm{log}\Lambda)$-plane as
$$ \log(\alpha) = \log \left( \frac{M_{\mathrm{Pl}}}{L^2 \rho_0} \right) + \frac{1}{n+1} \log \left( \frac{n M_{\mathrm{Pl}}}{\rho_0} \right) + \frac{n+4}{n+1} \log(\Lambda) - \frac{n+2}{n+1} \log(\beta) \, . $$
In the remainder of this article, the \emph{hat} notation used to designate dimensionless variables is dropped.

\section{Numerical methods: solving nonlinear PDE on unbounded domains with the finite element method}
\label{sec:methods}

Generic partial differential equations cannot be solved analytically, yet one can resort to numerical techniques to find approximate solution. In particular, we are interested in numerical approximations of the Klein-Gordon equation that governs the chameleon field.

Let $(n, \alpha) \in \mathbb{N} \times \mathbb{R}$ be the two dimensionless parameters of our model and consider the following boundary value problem:
\begin{equation}
    \alpha \Delta \phi(\mathbf{x}) = \rho(\mathbf{x}) - \phi^{-(n+1)}(\mathbf{x}) \quad \text{with} \quad
    \begin{cases}
    \phi(\mathbf{x}) & \underset{\|\mathbf{x}\| \to +\infty}{\longrightarrow} \phi_{\mathrm{vac}}  \\[5pt]
    \rho(\mathbf{x}) & \underset{\|\mathbf{x}\| \to +\infty}{\longrightarrow} \rho_{\mathrm{vac}}
    \end{cases} \, ,
\label{eqn:model_pb}
\end{equation}
where $\mathbf{x} \in \mathbb{R}^3$. One may note at least two difficulties in this problem:
\begin{enumerate}
    \item The r.h.s. term $\phi^{-(n+1)}$ makes the PDE nonlinear. More specifically, this second order PDE is \textit{semilinear} as the coefficients of the terms involving the highest-order derivatives of the unknown $\phi$ depend only on $\mathbf{x}$, not on $\phi$ or its derivatives \cite{semilinear};
    \item The chameleon field's profile is only known infinitely far away from the sources (asymptotic boundary conditions). However, computers' memory being finite, it is obviously not possible to produce a mesh of infinite spatial extension. Consequently, one has to come up with an alternative for properly imposing the boundary conditions.
\end{enumerate}
This section describes the implementation of \emph{femtoscope}, from the very basics of the finite element method to the more advanced concepts for tackling the two aforementioned difficulties.

\subsection{Basic ideas behind FEM}
\label{subsec:fem-basics}

The Finite Element Method (FEM) is a general numerical method for solving PDEs together with a set a of constraints imposed on the domain's boundary referred to as boundary conditions. The main idea behind FEM is to mesh a continuous spatial domain into a finite set of non-overlapping subdomains \textemdash \ the finite elements \textemdash \ over which the problem takes a simpler form. One of the key contributions in the development of FEM comes from the analysis of aircraft structures in the 1950s \cite{first-FEM}, which is why it is often associated with elasticity and structural analysis problems in aeronautical engineering. Since then, the method has been widely adopted in many other engineering disciplines, including heat transfer, electromagnetism, acoustics, and fluid dynamics (see e.g. Ref~\cite{Liu2022} for an historical perspective). In this subsection, we outline the key ideas behind this method on a generic linear second-order PDE
\begin{equation}
    \mathrm{div}\left[\mathbf{C}(\mathbf{x}) \boldsymbol{\nabla}u \right] + \mathbf{b}(\mathbf{x}) \cdot \boldsymbol{\nabla}u + a(\mathbf{x}) u = f \iff \sum_{i,j=1}^d \frac{\partial}{\partial x_i} \left[ C_{ij}(\mathbf{x}) \frac{\partial u}{\partial x_j} \right] + \sum_{i=1}^d b_i(\mathbf{x}) \frac{\partial u}{\partial x_i} + a(\mathbf{x}) u = f \, ,
\label{eqn:demo_edp}
\end{equation}
where $\mathbf{x} \in \Omega$, a smooth bounded open set of $\mathbb{R}^d$ with $d \in \mathbb{N}^*$, $\mathbf{C}(\mathbf{x}) = \left[C_{ij}(\mathbf{x})\right]_{1\leq i,j \leq d} \in \mathbb{R}^{d \times d}$, $\mathbf{b}(\mathbf{x}) = \left[b_i(\mathbf{x})\right]_{1 \leq i \leq d} \in \mathbb{R}^d$ and $a(\mathbf{x}) \in \mathbb{R}$. For this problem to be well-posed, it is necessary to supplement this equation with boundary conditions, imposed at the border $\Gamma$ of $\Omega$. A possible choice is to partition the border into the disjoint union $\Gamma = \Gamma_{\mathrm{D}} \cup \Gamma_{\mathrm{N}}$ and set
\begin{equation}
    u = u_{\mathrm{D}} \ \text{on} \ \Gamma_{\mathrm{D}} \quad \text{and} \quad (\mathbf{C} \boldsymbol{\nabla}u) \cdot \mathbf{n} = g_{\mathrm{N}} \ \text{on} \ \Gamma_{\mathrm{N}} \, ,
\label{eqn:demo_bc}
\end{equation}
where $\mathbf{n}$ is the outward normal vector to $\Gamma_{\mathrm{N}}$. The former condition is referred to as \emph{Dirichlet} or \emph{essential} boundary condition while the latter is called a \emph{Neumann} boundary condition. For the sake of simplicity, we further assume that $u_{\mathrm{D}} \equiv 0$ in the following.

\subsubsection{Weak formulation of a partial differential equation}
\label{subsubsec:weak}

The first step of FEM consists in transforming the boundary value problem into the so-called \textit{weak form}. This is achieved by multiplying Eq.~(\ref{eqn:demo_edp}) by a test-function $v$ that belongs to a functional space $V$ (to be specified later on) and integrating the resulting equation over the whole space $\Omega$. Concretely, this leads to:
\begin{equation}
    - \int_{\Omega} [\mathbf{C}(\mathbf{x}) \boldsymbol{\nabla} u] \cdot \boldsymbol{\nabla} v \, \mathrm{d}\mathbf{x} + \int_{\Gamma_{\mathrm{N}}} g_{\mathrm{N}} v \, \mathrm{d}\gamma + \int_{\Omega} (\mathbf{b}(\mathbf{x}) \cdot \boldsymbol{\nabla}u) v \ \mathrm{d} \mathbf{x} + \int_{\Omega} a(\mathbf{x}) u v \, \mathrm{d} \mathbf{x} = \int_{\Omega} fv \, \mathrm{d} \mathbf{x} \, ,
\label{eqn:continuous_weak_eqn}
\end{equation}
where we have made use of the divergence theorem and further imposed $v \equiv 0$ on $\Gamma_{\mathrm{D}}$ (Dirichlet boundary conditions are included in the functional space $V$). Rearranging the terms leads to the variational formulation of the problem:
\begin{equation}
    \text{Find } u \in V \text{ such that for all } v \in V, \ a(u, v) = l(v) \, ,
\label{eqn:continuous_weak_form}
\end{equation}
where $a$ is a bilinear form on $V \times V$ and $l$ is a linear form on $V$. The question whether the weak problem is well-posed i.e. has a unique solution was studied by mathematicians up until the 70s. Famous results are the \textsc{Lax-Milgram} theorem (sufficient conditions for well-posedness) and \textsc{Inf-Sup} theory (sufficient and necessary conditions for well-posedness) \cite{well-posedness-results}.

\subsubsection{Look for a solution in a finite dimensional function space}
\label{subsubsec:Vh}

The second and last fundamental idea of the finite element method is to approximate the infinite dimensional space $V$ (in which we look for the solution) by a smaller, finite dimensional space $V^h$ that can fit into a computer's memory. Let $N \coloneqq \mathrm{dim}(V^h)$ and $(w_i)_{1 \leq i \leq N}$ be a basis of $V^h$. Then any function $\phi^h \in V^h$ may be decomposed equivocally as
\begin{equation}
    \phi^h = \sum_{i=1}^N \Phi_i w_i \ , \quad \Phi_i \in \mathbb{R} \, .
\label{eqn:basis-decomposition}
\end{equation}

In the weak form (\ref{eqn:continuous_weak_form}), testing against all $v \in V$ is now equivalent to testing against all basis functions, such that the discrete weak formulation reads
\begin{equation}
    \text{Find } \mathbf{U} \in \mathbb{R}^N \text{ such that for all } i \in \{1, \dots, N \}, \ \sum_{j=1}^N U_j a(w_j, w_i)  = l(w_i) \ ,
\label{eqn:discrete_weak_form}
\end{equation}
which is nothing but a linear system of unknown $\mathbf{U} = (U_1, \dots , U_N)^T$, with matrix $\mathbf{A} = \left( a(w_j, w_i) \right)_{1 \leq i, j \leq N}$ and r.h.s. vector $\mathbf{L} = \left( l(w_i) \right)_{1 \leq i \leq N}$.

The remaining ingredient of FEM is the mesh, which is composed of simple cells such as triangles in 2D or tetrahedra in 3D (see Fig. \ref{fig:mesh_reduction}). A very common choice for the basis functions is to employ Lagrange polynomials associated with a given node and whose support is restricted to cells sharing this specific node.

\subsection{Handling nonlinearity}
\label{subsec:nonlinearity}

\subsubsection{Nonlinear solver}
\label{subsubsec:nonlinear}

\begin{table}[b]
\caption{\label{tab:math-notations} Mathematical notations introduced in Sec.~\ref{subsec:nonlinearity}.}
\renewcommand{\arraystretch}{1.2}
    \centering
    \begin{tabular}{ *{6}{q{c}{\mylen}} }
    \toprule
    \multicolumn{6}{c}{\textbf{Functionals}} \\[3pt]
    \coltwo{l}{$f_v$} & \coltwo{c}{$V \longrightarrow \mathbb{R}$} & \coltwo{r}{continuous weak form} \\
    \coltwo{l}{$\Tilde{f}_{v, \phi}$} & \coltwo{c}{$V \longrightarrow \mathbb{R}$} & \coltwo{r}{continuous linearized weak form} \\
    \coltwo{l}{$F$} & \coltwo{c}{$\mathbb{R}^N \longrightarrow \mathbb{R}^N$} & \coltwo{r}{discrete weak form} \\
    \coltwo{l}{$\Tilde{F}_{\phi^h}$} & \coltwo{c}{$\mathbb{R}^N \longrightarrow \mathbb{R}^N$} & \coltwo{r}{discrete linearized weak form} \\ \midrule
    \coltwo{c}{\textbf{Functions}} & \coltwo{c}{\textbf{Vectors}} & \coltwo{c}{\textbf{Matrices}} \\[3pt]
    \multicolumn{2}{c}{$u, v, \phi \in V$}     & \multicolumn{2}{c}{\multirow{2}{*}{$\mathbf{P}, \mathbf{Q}_k , \mathbf{U} , \delta \mathbf{U} \in \mathbb{R}^N$}} & \multicolumn{2}{c}{\multirow{2}{*}{$\mathbf{A} , \mathbf{B}_k \in \mathbb{R}^{N \times N}$}} \\
\multicolumn{2}{c}{$u^h , \phi^h \in V^h$} & \multicolumn{2}{c}{} & \multicolumn{2}{r}{} \\ \midrule
\colthree{c}{\textbf{Residual vector}} & \colthree{c}{\textbf{Residual}} \\[3pt]
\colthree{c}{$F(\mathbf{U}) \in \mathbb{R}^N$} & \colthree{c}{$\| F(\mathbf{U}) \|_{2} \in \mathbb{R}_+$} \\
\bottomrule
    \end{tabular}
\end{table}

The general methodology presented above holds as long as the PDE is linear, which is not the case of the Klein-Gordon equation (\ref{eqn:model_pb}). To bring this issue out, let us apply the techniques described in the former section to Eq.~(\ref{eqn:model_pb}). The asymptotic boundary conditions are deliberately left out as they are not relevant yet. We thus consider a simpler problem
\begin{equation}
    \alpha \Delta \phi(\mathbf{x}) = \rho(\mathbf{x}) - \phi^{-(n+1)}(\mathbf{x}) , \ \forall \mathbf{x} \in \Omega \quad \text{with} \quad \phi = \phi_{\mathrm{D}} \ \text{on} \ \Gamma \coloneqq \partial \Omega \, ,
\label{eqn:model_pb_bounded}
\end{equation}
where some artificial Dirichlet boundary condition has been introduced. Now let $v \in V$, the weak equation reads
$$ - \alpha \int_{\Omega} \boldsymbol{\nabla} \phi \cdot \boldsymbol{\nabla} v \, \mathrm{d} \mathbf{x} + \int_{\Omega} \phi^{-(n+1)} v \, \mathrm{d} \mathbf{x} = \int_{\Omega} \rho v \, \mathrm{d} \mathbf{x} \, . $$
However, the basis decomposition (\ref{eqn:basis-decomposition}) fails to produce a linear system precisely because of the nonlinear term. One successful approach for dealing with such nonlinear PDEs is to resort to iterative methods such as the Picard's method (also known as \textit{fixed-point iteration}, \textit{successive substitution} or \textit{nonlinear Richardson iteration}) or Newton type schemes (see e.g. Ref.~\cite{HPL_fem}). There seem to be some confusions in the literature regarding the derivation of the implementation. Table \ref{tab:math-notations} wraps up most notations introduced down below while Appendix \ref{sec:linesearch} provides the full expressions of the functionals at stake. Here $\|\cdot\|_2$ denotes the usual Euclidean norm (also called $l^2$ norm or 2-norm) over $\mathbb{R}^N$. The first point to clarify is that the linearization process occurs before the discretization step (i.e. on the continuous weak form). Let $v \in V$ and $f_v : V \to \mathbb{R}$ be the functional such that
$$ \forall u \in V, \ f_v(u) \coloneqq \alpha \int_{\Omega} \boldsymbol{\nabla} u \cdot \boldsymbol{\nabla} v \, \mathrm{d} \mathbf{x} - \int_{\Omega} u^{-(n+1)} v \, \mathrm{d} \mathbf{x} + \int_{\Omega} \rho v \, \mathrm{d} \mathbf{x} \, . $$
We are interested in finding $u$ such that $f_v(u) = 0$, for all test functions $v$. The general procedure to obtain Newton's iteration step is to write the approximate solution at the $(k+1)^{\mathrm{th}}$ iteration as a small increment from the previous approximation, that is $u_{k+1} = u_k + \delta u_k$. Plugging this new iterate into the expression of $f_v$ yields:
\begin{flalign}
 &\!\begin{aligned}
f_v(u_{k+1}) &= f_v(u_k + \delta u_k) \\[5pt]
&= \alpha \int_{\Omega} \boldsymbol{\nabla} u_k \cdot \boldsymbol{\nabla} v \, \mathrm{d} \mathbf{x}  + \alpha \int_{\Omega} \boldsymbol{\nabla} \delta u_k \cdot \boldsymbol{\nabla} v \, \mathrm{d} \mathbf{x} - \int_{\Omega} (u_k + \delta u_k)^{-(n+1)} v \, \mathrm{d} \mathbf{x} + \int_{\Omega} \rho v \, \mathrm{d} \mathbf{x} \, .
\end{aligned}&
\label{eqn:F_u_k+1}
\end{flalign}
We can now Taylor expand the nonlinear term at first order
\begin{flalign*}
(u_k + \delta u_k)^{-(n+1)} & = u_k^{-(n+1)} - (n+1) u_k^{-(n+2)} \delta u_k + o(\delta u_k) \, .
\end{flalign*}
By dropping $o(\delta u_k)$ terms and substituting this expansion in Eq.~(\ref{eqn:F_u_k+1}), we obtain Newton's linearization:
\begin{flalign}
 &\!\begin{aligned}
    f_v(u_{k+1}) \simeq \Tilde{f}_{v, u_k}(u_{k+1}) &\coloneqq \alpha \int_{\Omega} \boldsymbol{\nabla} u_k \cdot \boldsymbol{\nabla} v \, \mathrm{d} \mathbf{x} + \alpha \int_{\Omega} \boldsymbol{\nabla} \delta u_k \cdot \boldsymbol{\nabla} v \, \mathrm{d} \mathbf{x} - \int_{\Omega} u_k^{-(n+1)} v \, \mathrm{d} \mathbf{x} \\[5pt]
    & + (n+1) \int_{\Omega} u_k^{-(n+2)} \delta u_k v \, \mathrm{d} \mathbf{x} + \int_{\Omega} \rho v \, \mathrm{d} \mathbf{x} \, ,
\end{aligned}&
\label{eqn:F_tilde_delta}
\end{flalign}
where $\Tilde{f}_{v, u_k}$ is the linearized version of the functional $f_v$ around $u_k$. The next iterate $u_{k+1}$ is obtained by first solving $\Tilde{f}_{v, u_k}(u_{k+1}) = 0$ for $\delta u_k$ and then applying the update $u_{k+1} = u_k + \delta u_k$. Note that at this stage (continuous weak formulation), one can reformulate the above without having to use the auxiliary unknown $\delta u_k$. Indeed, replacing $\delta u_k$ by $u_{k+1} - u_k$ in Eq.~(\ref{eqn:F_tilde_delta}) yields
\begin{flalign}
 &\!\begin{aligned}
    \Tilde{f}_{v, u_k}(u_{k+1}) = \alpha \int_{\Omega} \boldsymbol{\nabla} u_{k+1} \cdot \boldsymbol{\nabla} v \, \mathrm{d} \mathbf{x} & - (n+2) \int_{\Omega} u_k^{-(n+1)} v \, \mathrm{d} \mathbf{x} \\[5pt]
    & + (n+1) \int_{\Omega} u_k^{-(n+2)} u_{k+1} v \, \mathrm{d} \mathbf{x} + \int_{\Omega} \rho v \, \mathrm{d} \mathbf{x} \, ,
\end{aligned}&
\label{eqn:F_tilde_u_k+1}
\end{flalign}
so that both the indirect and direct approaches are strictly equivalent in their continuous form. Therefore, the distinction between ``Picard method" and ``Newton method" in Ref.~\cite{Briddon_2021} really boils down to the discretization step previously described in Sec.~\ref{subsubsec:Vh}.

\begin{figure*}
    \centering
    \includegraphics[width=1.0\textwidth]{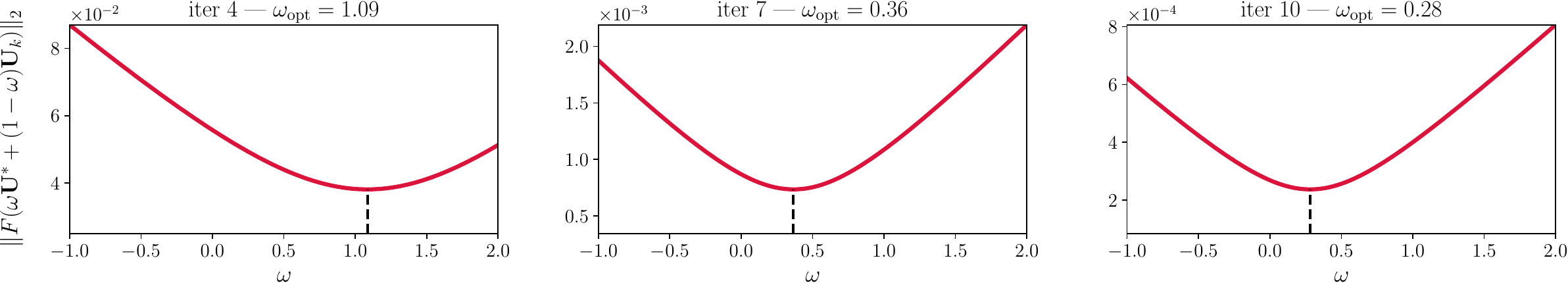}
    \caption{2-norm of the residual as a function of $\omega$ at the $\mathrm{4^{th}}$, $\mathrm{7^{th}}$ and $\mathrm{10^{th}}$ iterations of the Newton solver. The line-search algorithm consists in finding the value of $w$ such that $\omega \mathbf{U}^* + (1 - \omega) \mathbf{U}_k$ minimizes the residual at each iteration.}
    \label{fig:obj_func}
\end{figure*}

For the discretization step, it is useful to define the matrices and vectors at stake and distinguish between the ones that are iteration-dependent and the ones that are not.
\begin{flalign}
&\!\begin{aligned}
    \mathbf{A} \text{ such that } A_{ij} &\coloneqq \int_{\Omega} \boldsymbol{\nabla} w_i \cdot \boldsymbol{\nabla} w_j \, \mathrm{d} \mathbf{x} \\[5pt]
    \mathbf{B}_k \text{ such that } B_{ij}^k &\coloneqq \int_{\Omega} u_k^{-(n+2)} w_i w_j \, \mathrm{d} \mathbf{x} \\[5pt]
    \mathbf{P} \text{ such that } P_i &\coloneqq \int_{\Omega} \rho w_i \, \mathrm{d} \mathbf{x} \\[5pt]
    \mathbf{Q}_k \text{ such that } Q_i^k &\coloneqq \int_{\Omega} u_k^{-(n+1)} w_i \, \mathrm{d} \mathbf{x} \, .
\end{aligned}&
\end{flalign}
Note that matrix $\mathbf{B}_k$ and vector $\mathbf{Q}_k$ are not computed with the group finite element method (see e.g. Ref.~\cite{HPL_fem}). This technique consists in performing the following approximation:
\begin{equation*}
    \left( \phi^h \right)^{-(n+1)} = \left( \sum_{i=1}^N \phi_i w_i \right)^{-(n+1)} \simeq \sum_{i=1}^N \phi_i^{-(n+1)} w_i \, .
\label{eqn:group_fem}
\end{equation*}
Instead, \emph{femtoscope} first computes $\phi^h$ at the Gauss quadrature points and only then raises it to the power $(n+1)$. With these notations, $\mathbf{U}_{k+1}$ is solution of the linear system of unknown $\mathbf{U}^*$
$$\Tilde{F}_{u_k}(\mathbf{U}^*) = 0 \iff (\alpha \mathbf{A} + (n+1) \mathbf{B}_k) \mathbf{U}^* = (n+2) \mathbf{Q}_k - \mathbf{P} \, .$$
Because iterative techniques are sometimes subject to convergence issues, it is possible to introduce a so-called relaxation parameter $\omega \in ]0, 1]$ to prevent the new iterate from being ``too far away" from the former one. The update procedure is then a mere convex combination of $\mathbf{U}^*$ and $\mathbf{U}_{k}$ which reads:
$$ \mathbf{U}_{k+1} = \omega \mathbf{U}^* + (1 - \omega) \mathbf{U}_k \, . $$
The price to pay for this added stability to the algorithm is a potentially slower convergence. An efficient yet more costly approach is to employ a line search algorithm at each iteration, that is to find
$$ \omega_{\mathrm{opt}}^{k+1} \coloneqq \arg \min_{\omega} \| F(\omega \mathbf{U}^* + (1 - \omega) \mathbf{U}_k) \|_{2} \, . $$
Fig.~\ref{fig:obj_func} illustrates this procedure by representing the $L^2$-norm of the residual vector $F(\mathbf{U}_k)$ given by Eq.~(\ref{eqn:discrete-nonlinear}) as a function of $\omega$ for $k \in \{4, 7, 10 \}$. The convergence will also depend on the initial guess $\mathbf{U}_0$: the closer it is to the \textit{true} solution, the faster the convergence. There are no ready-made recipe for initializing the solution which is why it is often decisive to have insights into the physical problem to solve. In the case of the chameleon field, the minimum and maximum density values in the computational domain give bounds on the scalar field. For more specific cases where the field is to be studied in the vicinity of quasi-spherical objects, the analytical approximation developed in Ref.~\cite{khoury_weltman_original} and described later in section \ref{subsec:verif-case} is used in \emph{femtoscope} to initialize the field's DOFs.

\subsubsection{Stopping criteria}
\label{subsubsec:criteria}

\begin{figure}
    \centering
    \includegraphics[width=0.5\textwidth]{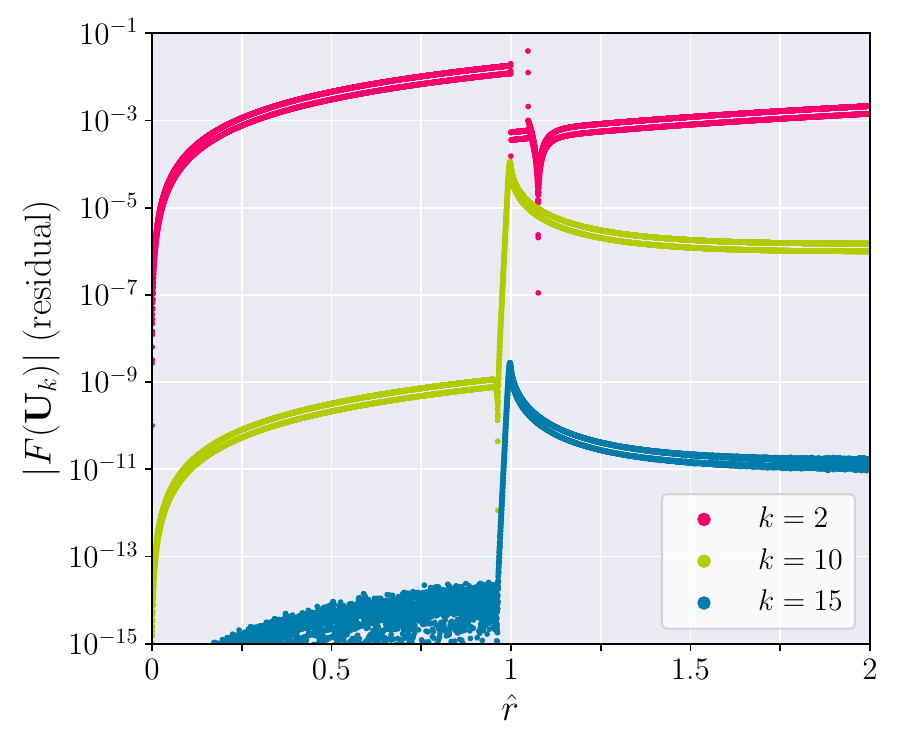}
    \caption{Residual as a monitoring tool. Absolute value of the residual vector at the $\mathrm{2^{nd}}$, $\mathrm{10^{th}}$ and $\mathrm{15^{th}}$ Newton iterations from 1D chameleon field computation with artificial Dirichlet boundary condition imposed at $\hat{r} = 5$. At $\hat{r}=1$, the density, which drives the chameleon field, drops down by five orders of magnitude. As a result, the residual tends to be large in this localized zone where the field undergoes rapid variations. The residual is efficiently reduced thanks to our Newton implementation.}
\label{fig:res_iter}
\end{figure}

The iterative algorithm must be terminated at some point. The relevant stopping criteria should be chosen such that they can quantitatively assess the convergence. To that extent, \emph{femtoscope} implements:

\begin{itemize}
    \item[--] a relative change condition, that is $ \|(\mathbf{U}_{k+1}- \mathbf{U}_k)/\mathbf{U}_k \|_2 \overset{?}{<} \delta_1 $. In other words, the algorithm terminates if the solution does not change significantly between two consecutive iterations;
    \item[--] a \textit{residual evaluation}. This is another very meaningful criterion regarding convergence. It simply consists in evaluating the nonlinearized functional $F$ at the current iteration and checking how close the computed value is to zero, that is $\|F(\mathbf{U}_k)\| \overset{?}{<} \delta_2$ for a given norm $\| \cdot \|$. It has the drawback of being an absolute criterion, which means $\delta_2$ is actually problem dependent. Fig.~\ref{fig:res_iter} shows the residual vector computed at the first three iterations of a 1D chameleon field FEM simulation (radial dependence only);
    \item[--] a maximum number of iterations.
\end{itemize}
We emphasize the fact that, beyond assessing convergence, the residual implementation is also a way to ascertain whether the computed solution weakly satisfies the original PDE. This is crucial since, as most nonlinear PDEs, the Klein-Gordon equation (\ref{eqn:model_pb}) has no closed-form solutions, making it difficult to validate the nonlinear solver implementation. The relative size of the Newton step is a very common way to assess convergence as well as it is independent of the typical scale of the variables. Hence, $\delta_1$ can be chosen virtually as small as desired, although it should be bigger than machine epsilon in practice. Empirically, we set $\delta_1 = 10^{-6}$.

\subsection{Handling unbounded coefficients with weight regularization}
\label{subsec:weights}

This subsection is an introduction to the next one as we will have to deal with PDEs exhibiting unbounded coefficients. Consider the generic second-order PDE (\ref{eqn:demo_edp}) and assume that some of its coefficients (namely $\mathbf{C}$, $\mathbf{b}$, $a$) or r.h.s. term are not bounded on $\Omega$. We do not make any particular assumption regarding where such singularities arise: they could be confined to the boundary or lie within the interior of the domain. This consequently puts constraints on the function space $V$ for the weak equation (\ref{eqn:continuous_weak_eqn}) to be well-defined.

When dealing with non-degenerate PDEs, Sobolev spaces are a suitable functional framework for analysis. However, in the presence of unbounded coefficients, a natural approach is to look for solutions in weighted Sobolev spaces (Ref.~\cite{cavalheiro_weighted_2008} reviews some results obtained in the study of such function spaces). From a mathematical viewpoint, the choice of an adequate weighted Sobolev space can make each integral of (\ref{eqn:continuous_weak_eqn}) well-defined and make the variational formulation (\ref{eqn:continuous_weak_form}) well-posed (\textit{existence} and \textit{uniqueness} of the solution). Nevertheless, in the actual FEM computation as partially discussed in section \ref{subsec:fem-basics}, there is no obvious lever for ensuring that the problem is numerically free of singularities. Yet, two solutions can be put forth:
\begin{itemize}
    \item[--] choose $V^h$ (the discrete counterpart of $V$) as a subspace of the adequate weighted Sobolev space together with basis functions $w_i$. Although it is arguably the most logical thing to do following mathematical proofs, having problem-dependent basis functions is not desirable, all the more as we are using an existing FEM library \textemdash\textit {Sfepy} \textemdash \ which we do not wish to modify internally;
    \item[--] regularize the PDE before deriving its weak form, i.e. modify the strong form (\ref{eqn:demo_edp}). This way, no changes to the FEM solver are required.
\end{itemize}
The latter idea leads to the introduction of a function $\varpi : \Omega \to \mathbb{R}^*$ with appropriate regularity that serves to weight Eq.~(\ref{eqn:demo_edp}):
\begin{flalign*}
   & \mathrm{div}\left[ \mathbf{C}(\mathbf{x}) \boldsymbol{\nabla}u \right] + \mathbf{b}(\mathbf{x}) \cdot \boldsymbol{\nabla}u + a(\mathbf{x}) u = f \\
   \iff & \varpi(\mathbf{x}) \mathrm{div}\left[ \mathbf{C}(\mathbf{x}) \boldsymbol{\nabla}u \right] + \varpi(\mathbf{x}) \mathbf{b}(\mathbf{x}) \cdot \boldsymbol{\nabla}u + \varpi(\mathbf{x}) a(\mathbf{x}) u = \varpi(\mathbf{x}) f \\
   \iff & \mathrm{div}\left[ \varpi(\mathbf{x}) \mathbf{C}(\mathbf{x}) \boldsymbol{\nabla}u \right] + \left[\varpi(\mathbf{x}) \mathbf{b}(\mathbf{x}) - \mathbf{C}(\mathbf{x})^T \boldsymbol{\nabla} \varpi (\mathbf{x}) \right] \cdot \boldsymbol{\nabla}u + \varpi(\mathbf{x}) a(\mathbf{x}) u = \varpi(\mathbf{x}) f \, .
\end{flalign*}
From there, one can choose a weight $\varpi$ such that the underlying weak formulation is well-posed \cite{Boulmezaoud_2, oh_1, weights_galerkin}.

\subsection{Handling asymptotic boundary conditions}
\label{subsec:asymptotic-bc-num}

This is the last step of the numerical implementation to be covered, building on the previous subsections. As highlighted in Ref.~\cite{martin_2019_PRD}, it is not possible to enforce boundary conditions at infinity straight away due to the finite extent of computational memory. The easiest workaround is truncation which consists in replacing the unbounded domain by a \textit{sufficiently large} bounded domain and applying the set of boundary conditions at the artificial border. In the context of the chameleon field, \textit{sufficiently large} would mean several times the maximum Compton wavelength in the domain. This method has at least two disadvantages: first, the domain has to be large which translates into a rather large linear system to solve \cite{truncation_goldstein}; and second, setting infinity at a finite distance can result in inaccurate solutions \cite{martin_2019_PRD}. A more interesting approach would be to derive a new set of exact boundary conditions at the artificial border as done in Refs.~\cite{truncation_1, truncation_2, truncation_3, truncation_4, truncation_5, truncation_6} but it has not been further investigated in our work.

A successful approach was implemented in Ref.~\cite{martin_2019_PRD} for solving the Klein-Gordon equation with asymptotic conditions via a \textit{shooting} technique. However, it is limited to one-dimensional cases and hinges on some symmetry in the density profile, which makes it hardly generalizable to higher dimensional cases with arbitrary matter distribution.

The broad approach employed in the following consists in mapping unbounded domains to bounded ones using appropriate transformation, thereby avoiding the introduction of an artificial outer boundary. Although appealing, this approach carries its own issues: as recalled in Ref.~\cite{truncation_2}, the mapping of an infinite domain into a finite one cannot be bounded. \textit{Ergo}, the new mapped problem will necessarily contain singularities in its finite domain. This issue is dealt with using the weight regularization method described in Sec.~\ref{subsec:weights}. The techniques described below were applied to a simple 3D unbounded Poisson problem governing the gravitational potential of a single body:
\begin{equation}
    \begin{cases}
    \Delta \Phi (\mathbf{x}) =
    \begin{cases}
       \alpha \rho(\mathbf{x}) & \text{inside the body} \\
       0 & \text{outside the body}
    \end{cases}
    \\[15pt]
    \Phi \underset{\|\mathbf{x}\| \to +\infty}{\longrightarrow} 0
    \end{cases} \, ,
\label{eqn:poisson_pot}
\end{equation}
whose analytical solution is known for perfect solid spheres and flat ellipsoids of revolution \textemdash \ see e.g. Refs.~\cite{maclaurin, milan11}. These techniques are illustrated in Fig.~\ref{fig:kelvin_transform}-\ref{fig:curves_int_ext} and compared in Fig.~\ref{fig:convergence_unbounded}.

\subsubsection{Compactification of the whole domain}
\label{subsubsec:compactification}

Here, the idea is to apply a global coordinates transformation $T : \Omega \to \Tilde{\Omega}$ such that $\Tilde{\Omega}$ is bounded, hence the term \textit{compactification}. Typically, \textit{tangent} or \textit{inverse hyperbolic tangent} functions are suitable transformations. To the best of our knowledge, this was first proposed by Ref.~\cite{first_compactification}. In Ref.~\cite{Zenginoglu_hyperboloidal_2011}, the author applies 1D compactification $x \in \mathbb{R}^+ \mapsto x/(1+x)$ (algebraic map) to solve hyperbolic PDEs on unbounded domains, while Ref.~\cite{tatiana_2016} uses \textit{logarithmic} mappings for advection-diffusion equations. Here, we wish to provide an insightful example in the framework of the finite element method, which is not considered in the previous references. Let us consider the Poisson problem (\ref{eqn:poisson_pot}) expressed in spherical coordinates $(r, \theta, \varphi)$ for a rotationally symmetrical body. Poisson equation then reads (see Appendix~\ref{eqn:spherical_laplacian} for the expression of the Laplacian)
\begin{equation}
    \frac{1}{r^2} \frac{\partial}{\partial r} \left( r^2 \frac{\partial \Phi}{\partial r} \right) + \frac{1}{r^2 \sin (\theta)} \frac{\partial}{\partial \theta} \left( \sin (\theta) \frac{\partial \Phi}{\partial \theta} \right) = \alpha \rho(r,\theta) \, .
\end{equation}
Let $\psi > 0$ be a scale factor and define the algebraic invertible compactification transform:
\begin{equation}
\begin{aligned}[t]
  T_r \colon \mathbb{R}^+ &\to [0, \psi[\\
  r &\mapsto \frac{\psi r}{1+r}
\end{aligned}
\qquad\text{;}\qquad
\begin{aligned}[t]
  T_r^{-1} \colon [0, \psi[ &\rightarrow \mathbb{R}^+\\
    \eta &\mapsto \frac{\eta}{\psi - \eta}
\end{aligned} \, .
\label{eqn:compact-transform}
\end{equation}
Following Sec.~\ref{subsec:weights}, let $\varpi(r)$ be a weight function to be defined explicitly later on and $v$ a test function. The weak equation then reads:
\begin{flalign}
\label{eqn:weak-uncompactified}
&\!\begin{aligned}
    \int_{\Omega} \varpi(r) \left[ \begin{pmatrix}
    r^2 \sin(\theta) & 0 \\ 0 & \sin(\theta)
    \end{pmatrix} \boldsymbol{\nabla}\Phi \right] \cdot \boldsymbol{\nabla}v \, \mathrm{d}r \mathrm{d}\theta &+ \int_{\Omega} r^2 \sin(\theta) \varpi'(r) \frac{\partial \Phi}{\partial r} v \, \mathrm{d}r \mathrm{d}\theta \\[5pt]
    & + \alpha \int_{\Omega} r^2 \sin(\theta) \varpi(r) \rho(r, \theta) v \, \mathrm{d}r \mathrm{d}\theta = 0 \, .
\end{aligned} &
\end{flalign}
In this expression, $\boldsymbol{\nabla} = (\partial_r, \partial_{\theta})^T$ instead of the usual gradient in polar coordinates. From there, we can apply the radial coordinate change $r \mapsto \eta$ leading to:
\begin{flalign}
\label{eqn:weak-compactified}
&\!\begin{aligned}
    \int_{\Tilde{\Omega}} \varpi(\eta) \left[ \begin{pmatrix}
    \dfrac{\eta^2}{\psi} \sin(\theta) & 0 \\ 0 & \dfrac{\sin(\theta) \psi}{(\psi - \eta)^2}
    \end{pmatrix} \boldsymbol{\nabla}\Phi \right] \cdot & \boldsymbol{\nabla}v \, \mathrm{d}\eta \mathrm{d}\theta + \int_{\Tilde{\Omega}} \frac{\eta^2}{\psi} \sin(\theta) \varpi'(\eta) \frac{\partial \Phi}{\partial \eta} v \, \mathrm{d}\eta \mathrm{d}\theta \\[5pt]
    & + \alpha \int_{\Tilde{\Omega}} \frac{\psi \eta^2}{(\psi - \eta)^4} \sin(\theta) \varpi(\eta) \rho(\eta, \theta) v \, \mathrm{d}\eta \mathrm{d}\theta = 0 \, .
\end{aligned}&
\end{flalign}
Now in order to remove the singularity when $\eta \to \psi$, we set $\varpi(\eta) \coloneqq (\psi - \eta)^2$. A few points should be noted:
\begin{itemize}
    \item[--] the last integral in Eq.~(\ref{eqn:weak-compactified}) is not singular when $\eta \to \psi$ because $\rho$ has a compact support;
    \item[--] the weight function does not have to be expressed in the $r$ variable;
    \item[--] the asymptotic condition on the gradient of the unknown field (that was discussed earlier on in Sec.~\ref{subsec:asymptotic-bc-phys} for the strong form of the PDE) is encompassed in the function space $V$ for the integrals of Eq.~(\ref{eqn:weak-uncompactified}) to be well-defined.
\end{itemize}
With this particular choice, the compactified and regularized weak form finally reads:
\begin{flalign}
&\!\begin{aligned}
    & \int_{\Tilde{\Omega}} \left[ \begin{pmatrix}
    \dfrac{\left[\eta(\psi-\eta)\right]^2}{\psi} \sin(\theta) & 0 \\ 0 & \psi \sin(\theta)
    \end{pmatrix} \boldsymbol{\nabla}\Phi \right] \cdot \boldsymbol{\nabla}v \, \mathrm{d}\eta \mathrm{d}\theta \\[5pt]
    & - 2 \int_{\Tilde{\Omega}} \frac{\eta^2 (\psi-\eta)}{\psi} \sin(\theta) \frac{\partial \Phi}{\partial \eta} v \, \mathrm{d}\eta \mathrm{d}\theta \\[5pt]
    & + \alpha \int_{\Tilde{\Omega}} \psi \left(\frac{ \eta}{\psi - \eta}\right)^2 \sin(\theta) \rho(\eta, \theta) v \, \mathrm{d}\eta \mathrm{d}\theta = 0 \, .
\end{aligned}&
\end{flalign}
This modified weak form together with homogeneous Dirichlet boundary condition at $\eta = \psi$ leads to a well-posed problem whose FEM solution is in excellent agreement with the analytical solution, see Fig.~\ref{subfig:conv-polar}.

\subsubsection{Domain splitting and Kelvin inversion technique}
\label{subsubsec:kelvin}

\begin{figure*}[t]
  \centering
  \includegraphics[width=0.95\textwidth]{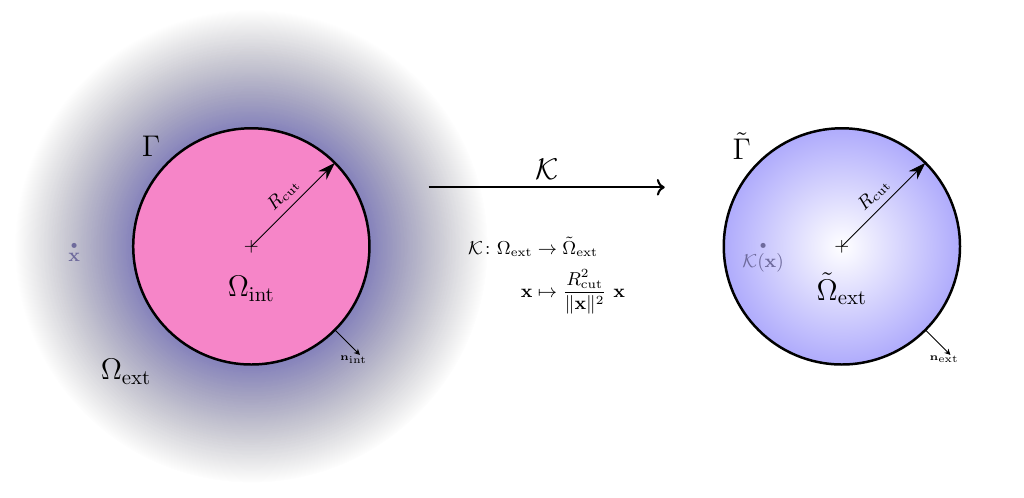}
  \caption{Illustration of the Kelvin transform; see the main text for notations and definitions.}
  \label{fig:kelvin_transform}
\end{figure*}

The above compactification transform (Eq.~\ref{eqn:compact-transform}) is applied to the whole domain, thus affecting the matter distribution. Mesh refinement nearby large density fluctuations or jumps is then more complicated as the domain to be meshed $\Tilde{\Omega}$ has already undergone a transformation with respect to the physical domain. In this regard, another fruitful approach is to split the unbounded domain into a bounded one, containing the sources of physical phenomena, and an unbounded one such that $\overline{\Omega} = \overline{\Omega}_{\mathrm{int}} \cup \overline{\Omega}_{\mathrm{ext}}$\footnote{Here, $\overline{\Omega}$ denotes the closure of the set $\Omega$.}. From there, several techniques can be found in the literature as early as the late 1970s: infinite elements introduced by Refs.~\cite{infinite_element_1, infinite_element_2, infinite_element_3} and implemented in the proprietary software \textsc{comsol}, combination of FEM and BEM (Boundary Element Method) \cite{infinite_bem_fem}, and several methods building on the Kelvin inversion \cite{kelvin_1, kelvin_2, oh_1, oh_2} or similar transformations \cite{Boulmezaoud_1, benjemaa_2019}.

\emph{femtoscope} implements two techniques based on the Kelvin inversion discussed in the following. For both, we first let $R_{\mathrm{cut}} > 0$ be the radius of a d-dimensional ball,  $d \in \{1, 2, 3\}$, defining the interior domain $\Omega_{\mathrm{int}}$ (big enough to encapsulate the various sources of physical interest) which is a bounded open subset of $\Omega$. Then the exterior domain $\Omega_{\mathrm{ext}} \coloneqq \Omega \setminus \overline{\Omega}_{\mathrm{int}}$ is mapped to a sphere of radius $R_{\mathrm{cut}}$ thanks to the Kelvin transform
\begin{equation}
    \begin{aligned}[t]
  \mathcal{K} \colon \Omega_{\mathrm{ext}} &\to \Tilde{\Omega}_{\mathrm{ext}}\\
  \mathbf{x} &\mapsto \frac{R_{\mathrm{cut}}^2}{\| \mathbf{x} \|^2} \ \mathbf{x}
\end{aligned} \, ,
\end{equation}
where $\Tilde{\Omega}_{\mathrm{ext}}$ denotes the image of the exterior domain through $\mathcal{K}$, also called the \textit{inversed exterior} from now on. Note that this mapping is not singular because by construction $\mathbf{0} \notin \Omega_{\mathrm{ext}}$. We further set $\Gamma \coloneqq \partial \Omega_{\mathrm{int}} \cap \partial \Omega_{\mathrm{ext}}$ the boundary delimiting the interior and exterior domains. This boundary is invariant under the Kelvin transform, that is $\mathcal{K}(\Gamma) \coloneqq \Tilde{\Gamma} = \Gamma$. Letting $d = \mathrm{dim}(\Omega)$, we have $\mathrm{dim}(\Gamma) = d-1$ so that $\Gamma$ has a measure of zero with respect to the Lebesgue measure on $\mathbb{R}^d$, hence $\int_{\Omega} = \int_{\Omega_{\mathrm{int}}} + \int_{\Omega_{\mathrm{ext}}}$. The Kelvin inversion can then be applied on integrals on $\Omega_{\mathrm{ext}}$, which is illustrated in Fig.~\ref{fig:kelvin_transform}. Once the solution is known on both domains, we can apply $\mathcal{K}^{-1}$ on $\Tilde{\Omega}_{\text{ext}}$ to reconstruct the solution on $\Omega$. \\

\paragraph{Virtual DOF connection at the shared frontier.}

\begin{figure*}[t]
    \centering
    \includegraphics[width=0.95\textwidth]{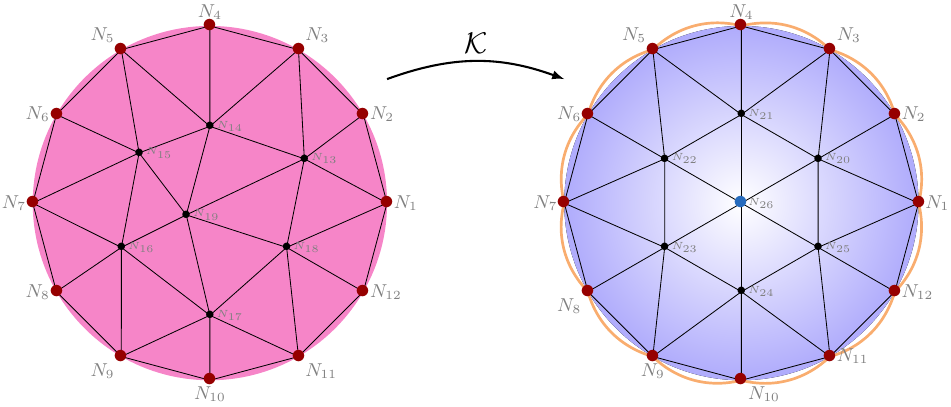}
    \caption{Interior and Inversed Exterior meshes generated for the \{domain splitting + Kelvin inversion + virtual connection of boundary DOFs between the two meshes\}. In computer's memory, nodes 1 to 12 are not duplicated, which results in a single FEM linear system after assembling the matrix of the linear system to be solved. The inversed exterior mesh must contain a node at the origin ($N_{26}$ in the figure) in order to impose the value of the unknown field at infinity. This requirement is relaxed for the interior mesh. Note that the Kelvin inversion does not preserve polygonal simplices, which are actually mapped to curved lines (in orange). Consequently, coefficients in the matrix involving these boundary DOFs are vitiated by a small error, which can be avoided using other types of inversion \cite{Boulmezaoud_1} or higher-order curved finite elements \cite{curved_elements}.}
    \label{fig:connected}
\end{figure*}

This first technique is based on the generation of two spherical meshes with matching facets at their border as illustrated in 2D in Fig. \ref{fig:connected}. In the actual FEM computation, these two sets of surface DOFs are merged into a unique set of DOFs (this is possible thanks to the definition of ``linear combination boundary conditions" in \textit{Sfepy}). This procedure results in a single linear system to be solved.

As in Sec.~\ref{subsubsec:compactification}, the mapping $\mathcal{K}$ leads to singular coefficients in the weak form near $\mathbf{0} \in \Tilde{\Omega}_{\mathrm{ext}}$. This ill-posed problem is regularized again using a weight function discussed above. Finding an appropriate weight is delicate in this case, because it cannot be applied on the exterior domain (which has to be regularized) independently of the interior domain (which is singularity-free). Thus, the weight must fulfill two requirements: (i) remove the singularity from the exterior domain and (ii) not introduce singularity in the interior domain as a side effect. In that respect, Sec.~\ref{subsubsec:compactification} provided a meaningful example for deriving relevant weights:
\begin{enumerate}
    \item write the weak formulation with an arbitrary weight function;
    \item apply the relevant coordinate transformation (compactification, Kelvin transform...) on the integrals;
    \item choose the weight function in the new coordinate system so as to remove the potential singularities.
\end{enumerate}

\paragraph{Iterative exchange of Dirichlet/Neumann boundary condition \textemdash \ the `ping-pong' technique.}

This idea takes its roots in domain decomposition methods developed in the field of High Performance Computing (HPC). HPC relies on the use of (massively) parallel architectures as well as huge global memory. Because several operations can be undertaken simultaneously, it is relevant to split the domain over which the PDE is defined into smaller sub-domains. Adjacent sub-domains must somehow exchange information at their common boundary, which requires substantial efforts and research in applied mathematics. In our case, we deal with two subdomains: the interior one $\Omega_{\mathrm{int}}$ and the inversed exterior one $\Tilde{\Omega}_{\mathrm{ext}}$. We here draw on a domain decomposition method called the \textit{Schwarz method without overlap} \cite{hpc_1, hpc_2} to connect the two domains. We now describe our algorithm.

Consider the two interrelated sub-problems at the $k^{\mathrm{th}}$ iteration, $k\geq1$:
\begin{equation}
    \begin{cases}
    \mathcal{L}(u_{\mathrm{int}}^{k+1}) = f &\text{ in } \Omega_{\mathrm{int}} \\[5pt]
    u_{\mathrm{int}}^{k+1} = u_{\mathrm{ext}}^{k} &\text{ on } \Gamma
    \end{cases}
    \quad \text{and} \quad
    \begin{cases}
    \mathcal{L}(u_{\mathrm{ext}}^{k+1}) = f &\text{ in } \Omega_{\mathrm{ext}} \\[5pt]
    u_{\mathrm{ext}}^{k+1} \longrightarrow u_{\infty} \in \mathbb{R} & \text{ at infinity} \\[5pt]
    \dfrac{\partial u_{\mathrm{ext}}^{k+1}}{\partial n_{\mathrm{ext}}} = -\dfrac{\partial u_{\mathrm{int}}^{k}}{\partial n_{\mathrm{int}}} & \text{ on } \Gamma
    \end{cases} \, ,
\label{eqn:ping-pong_principle}
\end{equation}
where $\mathcal{L}$ is an arbitrary linear differential operator. The algorithm then reads

\noindent \textbf{Initialization}: Solve the interior sub-problem with $u_{\mathrm{int}}^{1} = u_{\infty}$ on $\Gamma$ as an initial Dirichlet boundary condition.

\noindent $\mathbf{k^{\mathrm{th}}}$ \textbf{iteration}:
\begin{enumerate}
    \item Compute the flux of $u_\mathrm{int}^{k}$ through $\Gamma$ that is $\boldsymbol{\nabla} u_\mathrm{int}^{k} \cdot \mathbf{n}_{\mathrm{int}} = g_{_{|\Gamma}}$.
    \item In $\Omega_{\mathrm{ext}}$, apply the known asymptotic condition on the DOF(s) representing infinity.
    \item Because the two sub-domains share a common frontier $\Gamma$, the flux of the global unknown field $u$ from $\Omega_{\mathrm{int}}$ to $\Omega_{\mathrm{ext}}$ must be equal to the opposite of the flux of $u$ from $\Omega_{\mathrm{ext}}$ to $\Omega_{\mathrm{int}}$. We thus apply Neumann boundary condition using the ready-for-use $g_{_{|\Gamma}}$ function.
    \item Solve the exterior sub-problem, yielding $u_{\mathrm{ext}}^{k+1}$.
    \item Use the continuity condition of the global unknown $u$ at the frontier by retrieving the value of $u_{\mathrm{ext}}^{k+1}$ at $\Gamma$ and impose it as the new Dirichlet boundary condition of the interior sub-problem.
    \item Solve the interior sub-problem, yielding $u_{\mathrm{int}}^{k+1}$.
\end{enumerate}
The normal vectors $\mathbf{n}_{\mathrm{int}}$ and $\mathbf{n}_{\mathrm{ext}}$ are represented in Fig. \ref{fig:kelvin_transform}. In a nutshell, this algorithm boils down to alternatively computing \textit{Dirichlet-to-Neumann} (DtN) and \textit{Neumann-to-Dirichlet} (NtD) operators via the finite element method. The continuity of the solution across $\Gamma$ is imposed in the interior sub-problem while the continuity of its normal derivative is imposed in the exterior sub-problem. Ref.~\cite{ping-pong-proof} provides a convergence analysis for this iterative method in the framework of FEM. To the best of our knowledge, this approach has never been implemented in the literature for the specific purpose of solving linear PDEs on unbounded domains.

\begin{figure}
    \centering
    \includegraphics[width=0.8\textwidth]{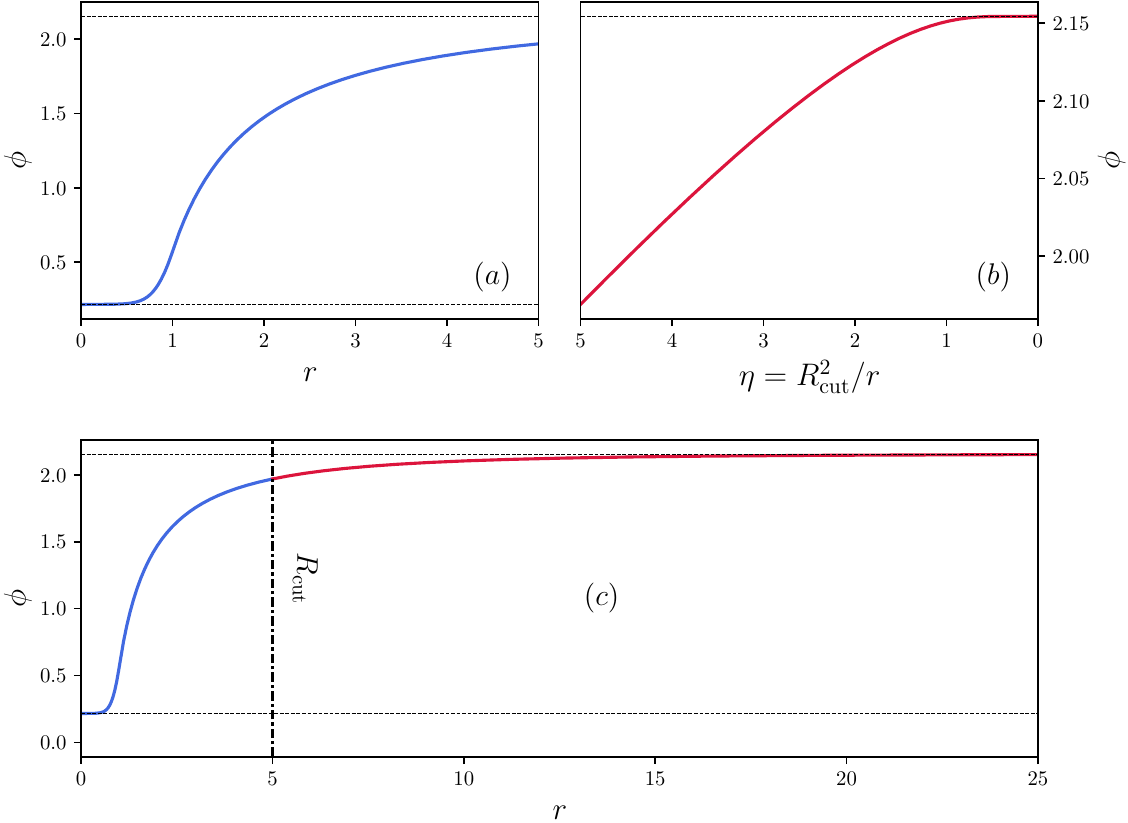}
    \caption{Assembling of the solution of Eq.~(\ref{eqn:model_pb}) on $\mathbb{R}_+$ (c) from the solution on the interior domain $r \in [0, R_{\mathrm{cut}}]$ (a) and on the inversed exterior domain $\eta \in [0, R_{\mathrm{cut}}]$ (b). This particular simulation showcases \emph{femtoscope} paramount feature: handling asymptotic conditions on the unknown. Indeed, there is no way to guess \textit{a priori} the value of the field at $R_{\mathrm{cut}}$ and imposing $\phi(R_{\mathrm{cut}}) = \phi_{\infty}$ would have led to a gross error. The dotted lines are set at $\phi_{\mathrm{min}} = \phi(r=0)$ and at $\phi_{\mathrm{max}} = \phi(r \to +\infty)$.}
\label{fig:curves_int_ext}
\end{figure}

One of the main assets of this technique compared to the virtual connection of DOFs described previously is that the two sub-domains are now completely separated. This implies that one can employ the weight regularization method on the inversed exterior domain without having to care about the interior one, giving more freedom in the choice of the relevant weight.

The two Kelvin-inversion-based techniques discussed above provide us with a numerical approximation of the solution on the interior domain $\Omega_{\mathrm{int}}$ and on the inversed exterior domain $\Tilde{\Omega}_{\mathrm{ext}}$. It is possible to reconstruct the solution on any bounded subset of the original domain $\Omega$ using $\mathcal{K}^{-1}$. This process is illustrated in Fig.~\ref{fig:curves_int_ext}.

\begin{figure}[t]
\subfloat[\label{subfig:conv-polar}]{%
  \includegraphics[height=7cm]{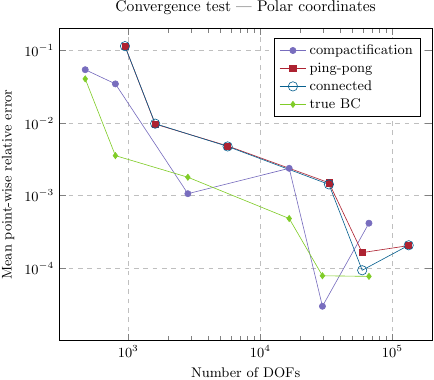}%
} \hspace{1cm}
\subfloat[\label{subfig:conv-cart}]{%
  \includegraphics[height=7cm]{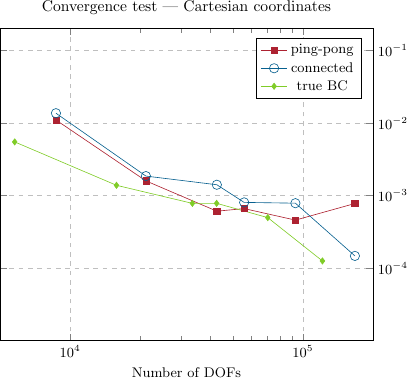}%
}
\caption{Convergence of the various algorithms for unbounded domains (discussed in Sec.~\ref{subsec:asymptotic-bc-num}) \textemdash \ (a) polar coordinates and (b) Cartesian coordinates. The test problem is a Poisson equation governing the gravitational potential of a flat ellipsoid. The mean point-wise relative error (y-axis) is computed by randomly sampling $n_s = 7091$ points in the interior domain over which the FEM-solution is compared against the analytical one. The \textit{true BC} curve (green) serves as another benchmark and is obtained by applying the exact Dirichlet boundary condition at the artificial border.}
\label{fig:convergence_unbounded}
\end{figure}

Finally, the three numerical techniques introduced in order to deal with PDEs defined on unbounded domains are compared in Fig.~\ref{fig:convergence_unbounded}, where we represent their respective convergence curve. The convergence rates obtained are compared to the result of standard FEM on a truncated domain with exact Dirichlet boundary condition at the artificial border (benchmark), depicted by the green curve. This curve shows how the error varies with respect to the number of DOFs in the \textit{ideal} case of standard FEM. We can see that the convergence rates of all three techniques are almost as good as the benchmark, regardless of the coordinate system used (Cartesian or polar). In the absence of rigorous mathematical proof, this convergence study validates our various implementations.

\section{\emph{femtoscope}}
\label{sec:femtoscope}

We now lay out how the numerical techniques presented above fit together into \emph{femtoscope}'s unified program. Our Python software is coded in an object-oriented fashion and builds upon an existing Python FEM package \textemdash \ \textit{Sfepy} \cite{sfepy}, which was early identified as a flexible open-source FEM library in Python. This code is being actively developed on GitHub\footnote{\url{https://github.com/sfepy/sfepy}. Last visited: June $\mathrm{1^{st}}$, 2022.}.

\begin{figure*}
    \centering
    \includegraphics[width=\textwidth]{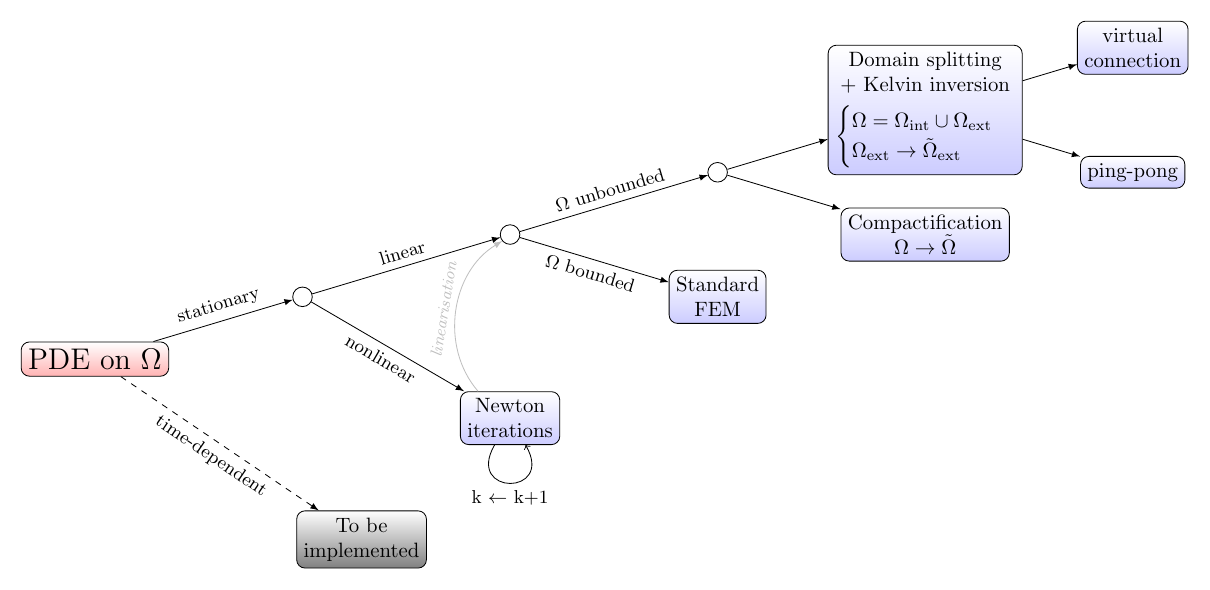}
    \caption{Overview of \emph{femtoscope} decision tree with respect to the nature of the PDE to be solved.}
    \label{fig:femtoscope_tree}
\end{figure*}

\emph{femtoscope} attempts to encompass the widest range of physical problems (e.g. linear one or not, defined on a bounded or unbounded spatial domain or not...). The handling of time-dependent problems is in progress. The decision tree for achieving this purpose is depicted in Fig.~\ref{fig:femtoscope_tree}.

Several test-cases were run in order to ascertain the  validity of the code. The techniques for dealing with asymptotic conditions were tested against the analytical solution to the problem of the gravitational potential inside and outside an oblate spheroid, governed by Eq.~(\ref{eqn:poisson_pot}). The solution to this problem inside the spheroid was found by Maclaurin \cite{maclaurin} while we referred to Ref.~\cite{milan11} for the solution outside the body using oblate spheroidal coordinates. The nonlinear custom Newton solver was first tested on bounded domains, where the chameleon field is supposedly known at the boundary. This configuration is that of an empty vacuum chamber (of density strictly over absolute zero) surrounded with thick walls. Then, in a certain part of the chameleon parameter space $(\beta, \Lambda, n)$, the walls would be screened so that deep inside them, the field would reach the value that minimizes the effective potential, which is known analytically (see Eq.~\ref{eqn:phi_vac}). In such cases, \emph{femtoscope} outputs were compared against SELCIE ones \cite{Briddon_2021}, whose code is accessible on GitHub\footnote{\url{https://github.com/C-Briddon/SELCIE}. Last visited: June $\mathrm{1^{st}}$, 2022.}. Appendix~\ref{sec:selcie_vs_femtoscope} provides a comparison between the two codes on a given test-case.

\subsection{Mesh generation}
\label{subsec:gmsh}

Meshes are created using the \textit{Gmsh} software \cite{gmsh}, a two- and three-dimensional finite element mesh generator with a built-in CAD\footnote{Computer-Aided Design} engine. Its Python API\footnote{Application Programming Interface} brings greater flexibility and allows us to automate the meshing of recurrent geometries. For an easier handling of matter distribution in the interior domain, shapes can be incorporated into a spherical background mesh and tagged with a unique identifier. As mentioned in Sec.~\ref{subsubsec:kelvin}, the interior and exterior meshes must have matching surface elements for using both the \textit{virtual connection method} or the \textit{ping-pong method}. All meshes are saved using the legacy Visualization Toolkit format (VTK).

\subsection{Coordinate systems and reduction of the space dimension in case of rotational invariance}
\label{subsec:coors}

\begin{figure}
    \centering
    \includegraphics[width=0.60\textwidth]{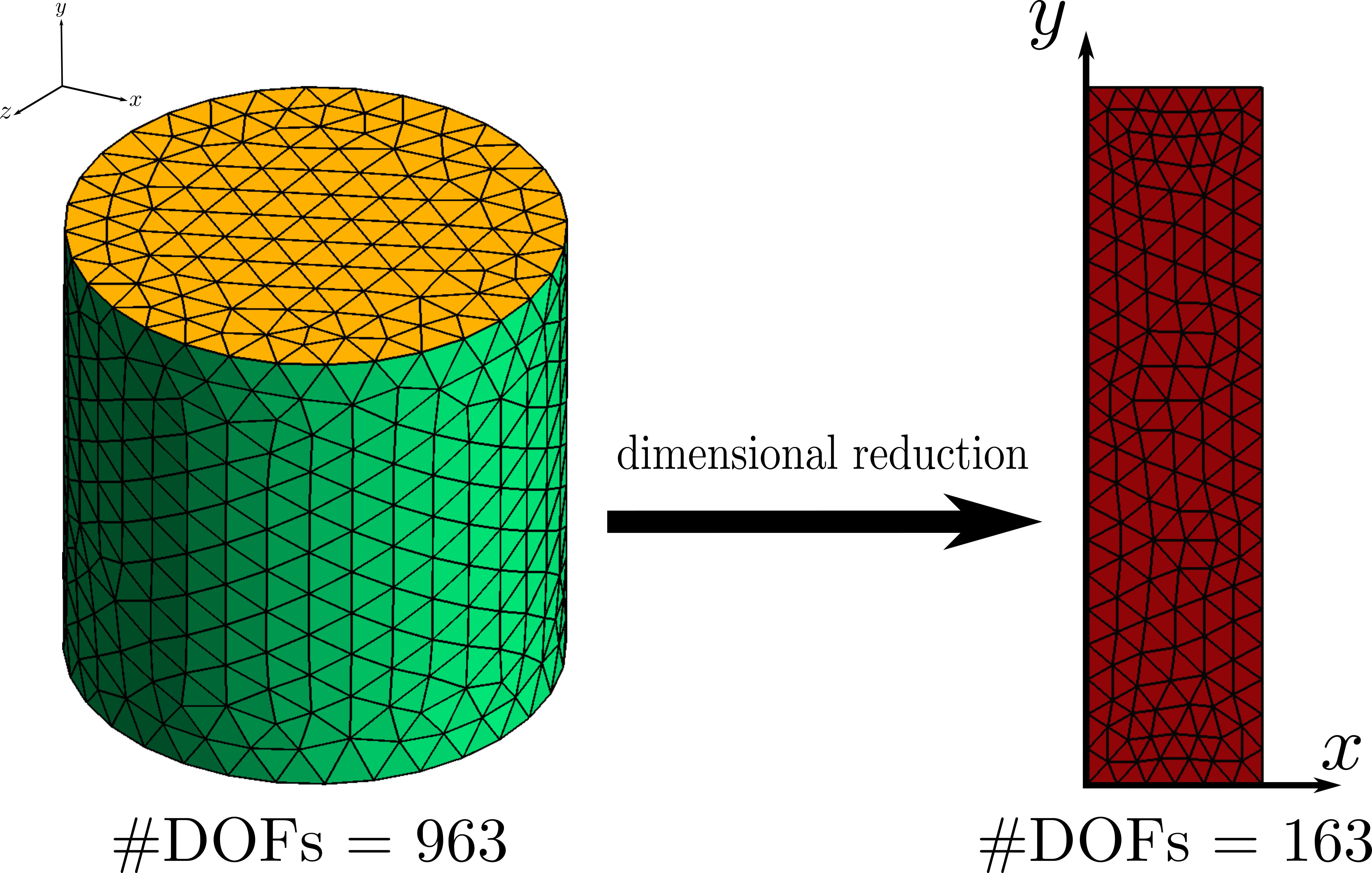}
    \caption{Illustration of the mesh size reduction process when dealing with axisymmetric setups (a cylinder here) in Cartesian coordinates.}
    \label{fig:mesh_reduction}
\end{figure}

\emph{femtoscope} implements polar coordinates and Cartesian coordinates for two-dimensional simulations and Cartesian coordinates for three-dimensional simulations. Reduction of the space dimension is only possible in certain symmetrical configurations, especially when the matter distribution is invariant by rotation around a given axis. In such cases, this axis is chosen such that $\partial/\partial \varphi \equiv 0$ in spherical coordinates Eq.~(\ref{eqn:spherical_laplacian}) or in cylindrical coordinates Eq.~(\ref{eqn:cylindrical_laplacian}). In the latter case, the $y$-axis coincides with the revolution axis and we introduce the Cartesian coordinates $(x, y) \in \mathbb{R}_+^* \times \mathbb{R}$ and consider the modified Laplacian
\begin{equation}
   \frac{1}{x} \frac{\partial}{\partial x} \left( x \frac{\partial f}{\partial x} \right) + \frac{\partial^2 f}{\partial y^2} \, .
\label{eqn:cartesian_laplacian_mod}
\end{equation}
Therefore, any 3D-axisymmetric setup can be worked out in a half plane. The unknown field is necessarily symmetric about the $y$-axis. This symmetry is accounted on the $x>0$ half-domain by imposing the Neumann condition
$$ \frac{\partial f}{\partial x}\bigg|_{x=0} = 0 \, .$$
This strategy to reduce the computational cost of a simulation is illustrated in Fig.~\ref{fig:mesh_reduction}. Two-dimensional FEM simulations should be preferred over three-dimensional ones whenever possible as the computational cost is greatly alleviated ($\sim 500$ times faster, empirically\footnote{This figure was obtained empirically by the authors. We first ran and timed a 3D simulation, then checked the error with respect to the analytical case, and finally ran and timed a 2D simulation that would exhibit the same error level.}).

\section{Chameleon field around the Earth}
\label{sec:chameleon-earth}

In the pursuit of being able to accurately model the chameleon $\mathrm{5^{th}}$-force in satellite-reachable regions of the Solar system, it is a good start to begin with the Earth environment. We first use \emph{femtoscope} on an homogeneous solid sphere immersed in a background medium of constant density. We then consider a more realistic physical model, still 1D, but with typical density profiles. In the remainder of the article, all simulations are conducted with the potential exponent set at $n=1$. Moreover, the dimensionless parameter $\alpha$ depends on the choice of the characteristic physical scales $L_0$ and $\rho_0$ (see Eq.~\ref{eqn:kg3}). Whenever relevant, we work with $L_0 = R_{\mathrm{Earth}} = 6371 \ \mathrm{km}$ and $\rho_0 = 1 \ \mathrm{kg . m^{-3}}$. 

\subsection{Perfect homogeneous solid sphere: verification of the implementation}
\label{subsec:verif-case}

We consider a perfect solid sphere of radius $R_A$ with constant density $\rho_{\mathrm{in}}$ surrounded by a medium of density $\rho_{\mathrm{vac}}$ assimilated to vacuum. This first application serves as a validation test-case since it is one of the simplest mass-distribution configurations. Indeed, the problem is reduced to a one-dimensional radial problem, for which approximated analytical solutions exist \cite{khoury_weltman_original}. For the dimensionless problem (\ref{eqn:model_pb}), the commonly used expression reads
\begin{equation}
    \phi(r) = \begin{cases}
        \rho_{\mathrm{in}}^{-\frac{1}{n+1}} & \text{if } r < R_{\mathrm{TS}} \, , \\[10pt]
        \rho_{\mathrm{in}}^{-\frac{1}{n+1}} + \dfrac{\rho_{\mathrm{in}}}{3 \alpha} \left( \dfrac{r^2}{2} + \dfrac{R_{\mathrm{TS}}^3}{r} - \dfrac{3}{2} R_{\mathrm{TS}}^2 \right) & \text{if } R_{\mathrm{TS}} \leq r \leq R_A \, , \\[10pt]
        \rho_{\mathrm{vac}}^{-\frac{1}{n+1}} - \dfrac{K}{r} \exp\left[-m_{\mathrm{vac}} (r-R_A)\right] & \text{if } r > R_A \, ,
    \end{cases}
\label{eqn:analytical_screened}
\end{equation}
in the screened regime, and
\begin{equation}
    \phi(r) = \begin{cases}
       \rho_{\mathrm{vac}}^{-\frac{1}{n+1}} - \dfrac{K}{R_A} + \dfrac{\rho_{\mathrm{in}}}{6 \alpha} (r^2 - R_A^2) & \text{if } r \leq R_A \, , \\[10pt]
       \rho_{\mathrm{vac}}^{-\frac{1}{n+1}} - \dfrac{K}{r} \exp[-m_{\mathrm{vac}} (r-R_A)] & \text{if } r > R_A \, ,
    \end{cases}
\label{eqn:analytical_unscreened}
\end{equation}
in the unscreened regime. In the above expressions,
\begin{itemize}
    \item[--] $m_{\mathrm{vac}} = \sqrt{ \dfrac{n+1}{\alpha} \rho_{\mathrm{vac}}^{\frac{n+2}{n+1}}}$ is the effective mass of the field in vacuum;
    \item[--] $K = \dfrac{\rho_{\mathrm{in}}}{3 \alpha} \left( R_A - \dfrac{R_{\mathrm{TS}}^3}{R_A^2} \right) \left( \dfrac{1}{R_A^2} + \dfrac{m_{\mathrm{vac}}}{R_A} \right)^{-1}$;
    \item[--] $R_{\mathrm{TS}}$ is the thin-shell radius. In the screened regime, it is computed as the only real root of the $3^{\mathrm{rd}}$-order polynomial of the variable $X$
    $$ \left[ - \frac{m_{\mathrm{vac}}}{3 \alpha (1 + m_{\mathrm{vac}} R_A)} \right] X^3 + \frac{1}{2\alpha} X^2 + \frac{1}{\rho_{\mathrm{in}}} \left( \rho_{\mathrm{vac}}^{-\frac{1}{n+1}} - \rho_{\mathrm{in}}^{-\frac{1}{n+1}} \right) - \frac{R_A^2}{3 \alpha} \left( \frac{1}{1 + m_{\mathrm{vac}} R_A} + \frac{1}{2} \right) $$
    that lies within the interval $[0, R_A]$. If this polynomial has no such root, it means that we are in the unscreened regime, and $R_{\mathrm{TS}} \equiv 0$.
\end{itemize}
This analytical approximation is useful from a phenomenological point of view to study the profile of the chameleon field around almost spherically symmetrical objects. In particular, it provides insights into the screening of an object.

\begin{figure}[t]
    \centering
    \includegraphics[width=0.6\textwidth]{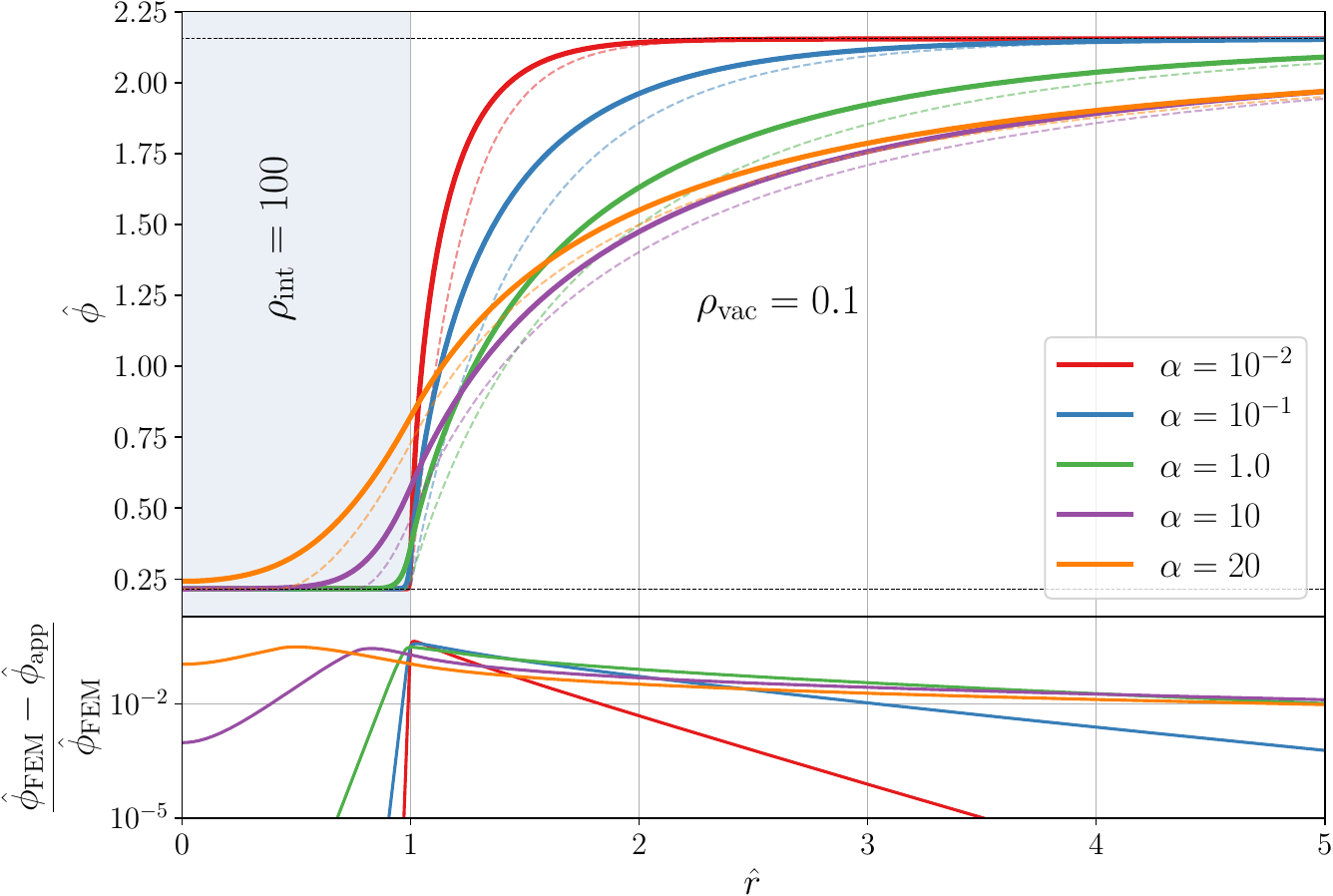}
    \caption{Radial profiles of the chameleon field for various values of $\alpha$, exhibiting the transition between the \textit{screened regime} ($\alpha \in \{ 10^{-2}, 10^{-1} \}$) and the \textit{unscreened regime} ($\alpha = 20$). The solid lines correspond to \emph{femtoscope}'s outputs $\hat{\phi}_{\mathrm{FEM}}$ while the dashed ones are the associated analytical approximations $\hat{\phi}_{\mathrm{app}}$ given by Eqs.~(\ref{eqn:analytical_screened}) and (\ref{eqn:analytical_unscreened}).}
    \label{fig:chameleon_sphere}
\end{figure}

Fig.~\ref{fig:chameleon_sphere} shows various chameleon field profiles obtained by FEM in polar coordinates (solid lines). The asymptotic condition is handled via the \textit{virtual DOF connection} method (see Sec.~\ref{subsubsec:kelvin}) with $R_{\mathrm{cut}} = 5$. This plot brings once again to the fore the importance of the FEM implementation on unbounded domains:
\begin{itemize}
    \item[--] in the screened regime, the field quickly reaches the value that minimizes the effective potential in vacuum $\phi_{\mathrm{vac}}$ so that one could apply this Dirichlet boundary condition at $R_{\mathrm{cut}}$ without making a big error;
    \item[--] for $\alpha \geq 1$ however, the field grows more slowly towards $\phi_{\mathrm{vac}}$ so that it would not have any physical sense to impose Dirichlet boundary condition at $R_{\mathrm{cut}}$.
\end{itemize}

In that same figure, we represent the analytical approximations in dashed lines. This approximation is not to be used if quantitative results are expected, even in the simplistic case of homogeneous perfect spheres. Indeed, as emphasized in Refs.~\cite{nfw_halos, martin_2019_PRD}, this analytical approximation is only valid in a certain region of the chameleon parameter space, which is why Eq.~(\ref{eqn:analytical_screened}) and (\ref{eqn:analytical_unscreened}) should not serve as a benchmark. In this respect, Table~\ref{tab:res-ana-num} clearly shows that the residual of the analytical approximation is larger by many orders of magnitude in 2-norm than that of the numerical solution after convergence. Besides, one may have noticed that there seem to be a relation between $\alpha$ and the size of the residual, especially the greater $\alpha$, the bigger the residual. This relation cannot be ascribed to a poor convergence of the Newton algorithm as the relative change of the numerical approximation between the last two iterations $\|(\mathbf{U}_{20}- \mathbf{U}_{19})/\mathbf{U}_{19} \|_2$ (see Sec.~\ref{subsubsec:criteria}) is consistently below $10^{-14}$ for all $\alpha$. A better explanation is linked to the fact that the residual, as defined in this article, is an \emph{absolute} quantity and not a \emph{relative} one (see Appendix~\ref{sec:alpha-res_relation}).

Appendix~\ref{sec:selcie_vs_femtoscope} provides a relevant comparison between \emph{femtoscope} and SELCIE on this specific test-case (see Fig.~\ref{fig:selcie_vs_femtoscope}). The overall excellent agreement between the two codes' outputs on a carefully chosen example brings further confidence regarding our implementation. This appendix further investigates the influence of the truncation radius $R_{\mathrm{cut}}$ on the accuracy of the solution.

\begin{table}
\renewcommand{\arraystretch}{1.2}
\caption{\label{tab:res-ana-num} 2-norm of the residual of Fig.~\ref{fig:chameleon_sphere} curves (20 iterations).}
\begin{ruledtabular}
\begin{tabular}{lcc}
& Residual analytical approximation & Residual numerical solution \\ \hline
$\alpha = 10^{-2}$ & $2.5 \times 10^{-2}$ & $9.7 \times 10^{-8}$\\
$\alpha = 10^{-1}$ & $4.0 \times 10^{-2}$ & $3.6 \times 10^{-8}$\\
$\alpha = 1$ & $6.0 \times 10^{-2}$ & $4.2 \times 10^{-6}$\\
$\alpha = 10$ & $8.1 \times 10^{-2}$ & $6.7 \times 10^{-5}$\\
$\alpha = 20$ & $1.1 \times 10^{-1}$ & $1.2 \times 10^{-4}$
\end{tabular}
\end{ruledtabular}
\end{table}



\subsection{Realistic Earth model}
\label{subsec:realistic-earth}

We now consider a more realistic treatment of chameleon gravity in the Earth vicinity. We look for quantitative values of the $\mathrm{5^{th}}$-force predicted by the chameleon model in Earth orbit. When relevant, the altitude is chosen to be that of the GRACE-FO satellites\footnote{\url{https://gracefo.jpl.nasa.gov/}. Last visited: September $1^{\mathrm{st}}$, 2022.}, i.e. around 500 km \cite{grace-fo}. As mentioned in Sec.~\ref{subsec:cham-physics}, the chameleon field alters geodesics. It is natural to study the effect of the underlying perturbing acceleration on geodesy satellites, hence the need to quantify it. To put things into perspective, the chameleon $\mathrm{5^{th}}$-force can be compared to other known physical effects taking place in orbit around the Earth. Especially, it is meaningful to compare the chameleonic force against the relativistic correction to Newtonian gravity and to Newtonian gravity itself. At first order, this correction reads
\begin{equation}
    \delta a_{\mathrm{GR}} = \frac{3}{r^3} \left( \frac{\mu_{\mathrm{Earth}}}{c} \right)^2 \simeq 10^{-9} a_{\mathrm{Newton}}
\label{eqn:delta_GR}
\end{equation}
for a circular orbit \cite{satellite_orbits}, where $\mu_{\mathrm{Earth}}$ is the Earth's standard gravitational parameter. This is already about 9 orders of magnitude smaller than Newtonian attraction for typical satellite altitudes (from low Earth orbits to the geostationary one).

\begin{figure}
    \centering
    \includegraphics[width=0.75\textwidth]{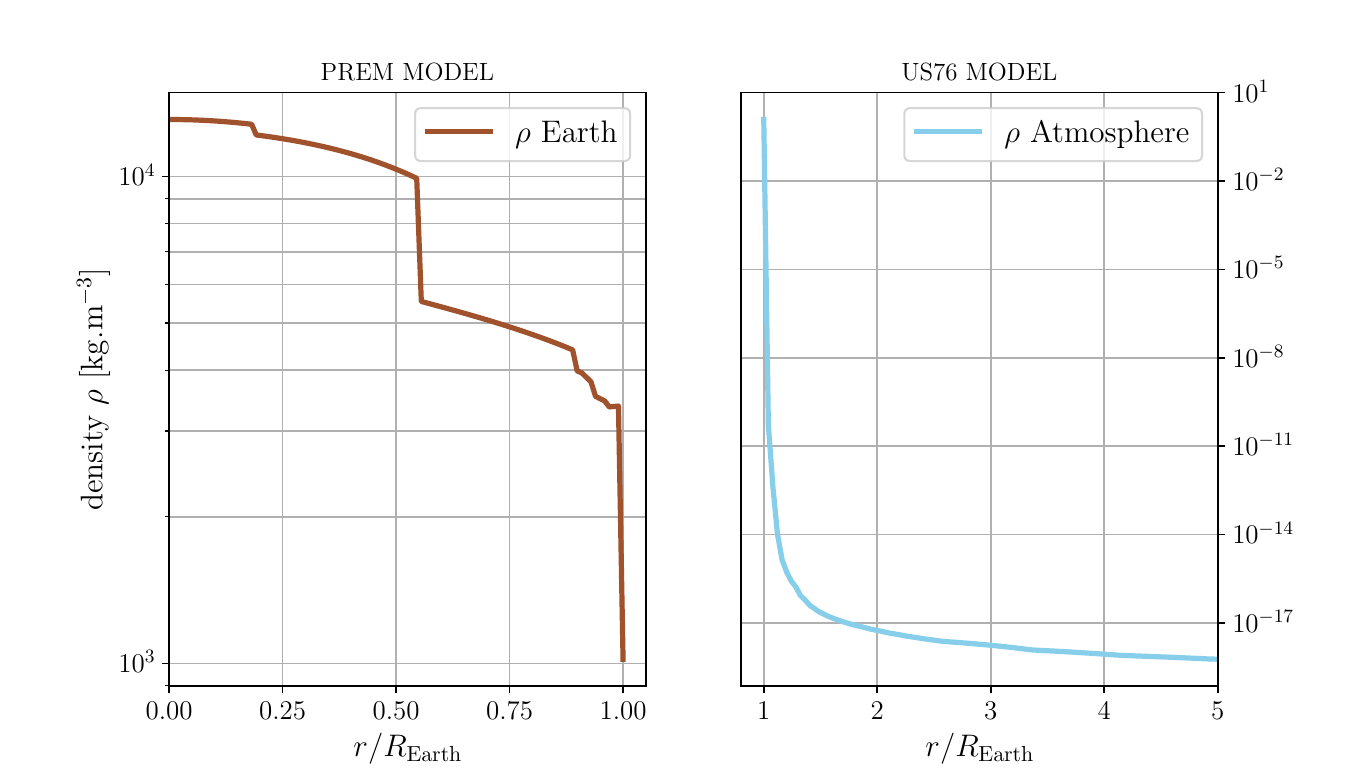}
    \caption{Implementation of realistic density models inside and outside the Earth for numerical simulations. Inside the Earth, the density is retrieved from the \emph{Preliminary reference Earth model} (PREM) \cite{PREM} while the atmospheric density is approximated by the US76 model\footnote{Data downloaded from \url{http://www.braeunig.us/space/atmos.htm}, (especially for the density between 1000 km - 36000 km altitude). Last visited: June $\mathrm{1^{st}}$, 2022.}. For altitudes higher than the geostationary altitude, we make the assumption that the density no longer decreases and stays at its minimum value $\rho_{\mathrm{vac}} = 4.04 \times 10^{-19} \  \mathrm{kg.m^{-3}}$.}
    \label{fig:dens_model_curves}
\end{figure}

\begin{figure}
    \centering
    \includegraphics[width=0.5\textwidth]{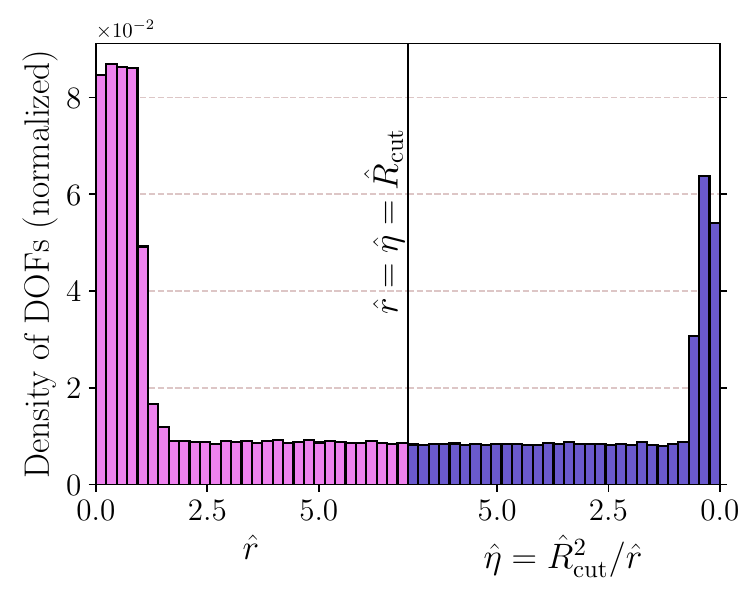}
    \caption{Distribution of DOFs of the realistic Earth 1D mesh. In the interior domain (left side), the mesh is refined around the various density jumps inside the Earth and at the transition between the crust and the atmosphere (see Fig.~\ref{fig:dens_model_curves}). In the inversed exterior domain (right side), the mesh is refined around the characteristic slope break at $\hat{\eta} = m_{\mathrm{vac}} \hat{R}_{\mathrm{cut}}^2 / 3$ which can be seen on Fig.~\ref{fig:curves_int_ext}(b). The \textit{hat} notation is used to denote dimensionless quantities: $\hat{r} = r/R_{\mathrm{Earth}}$, $\hat{R}_{\mathrm{cut}} = R_{\mathrm{cut}}/R_{\mathrm{Earth}} = 7$ and $\hat{\eta} = \hat{R}_{\mathrm{cut}}^2 / \hat{r}$.}
    \label{fig:distribution_DOFs}
\end{figure}

The first step is to implement a realistic model of the density inside and around the Earth as in Fig.~\ref{fig:dens_model_curves}. The use of purely radial models allows us to conduct numerical simulations in 1D, much cheaper than their 2D or 3D counterpart\footnote{\textit{De facto}, the Earth flattening at the poles cannot be taken into account despite being one of the major perturbing accelerations \cite{satellite_orbits}. Ref~\cite{burrage_ellipticity} shows that ellipsoidal departures from spherical symmetry results in an enhancement of the chameleon force.}. The density decreases from $1.3 \times 10^4 \ \mathrm{kg.m^{-3}}$ at the center of the Earth to barely $4.0 \times 10^{-19} \ \mathrm{kg.m^{-3}}$ beyond the geostationary altitude, which represent a variation over nearly 23 orders of magnitude. Moreover, it is subject to a 3 orders of magnitude jump at the interface between the Earth and the atmosphere. Density being the source of the field, the mesh employed in numerical simulations has to be very fine around such rapid variations (see Fig.~\ref{fig:distribution_DOFs}), and we set the relaxation parameter to $\omega = 0.5$ (experimentally determined to ensure convergence). The truncation radius $R_{\mathrm{cut}}$ is set at $7 R_{\mathrm{Earth}}$ because the density is assumed constant beyond this altitude. To check the relevance of such models, we solved for the Newtonian potential governed by Poisson equation (\ref{eqn:poisson_pot}) and found the conventional value of gravitational acceleration on Earth, $\|\mathbf{g}\|$, of about 9.8 $\mathrm{m.s^{-2}}$.

\begin{figure}
    \centering
    \includegraphics[width=0.5\textwidth]{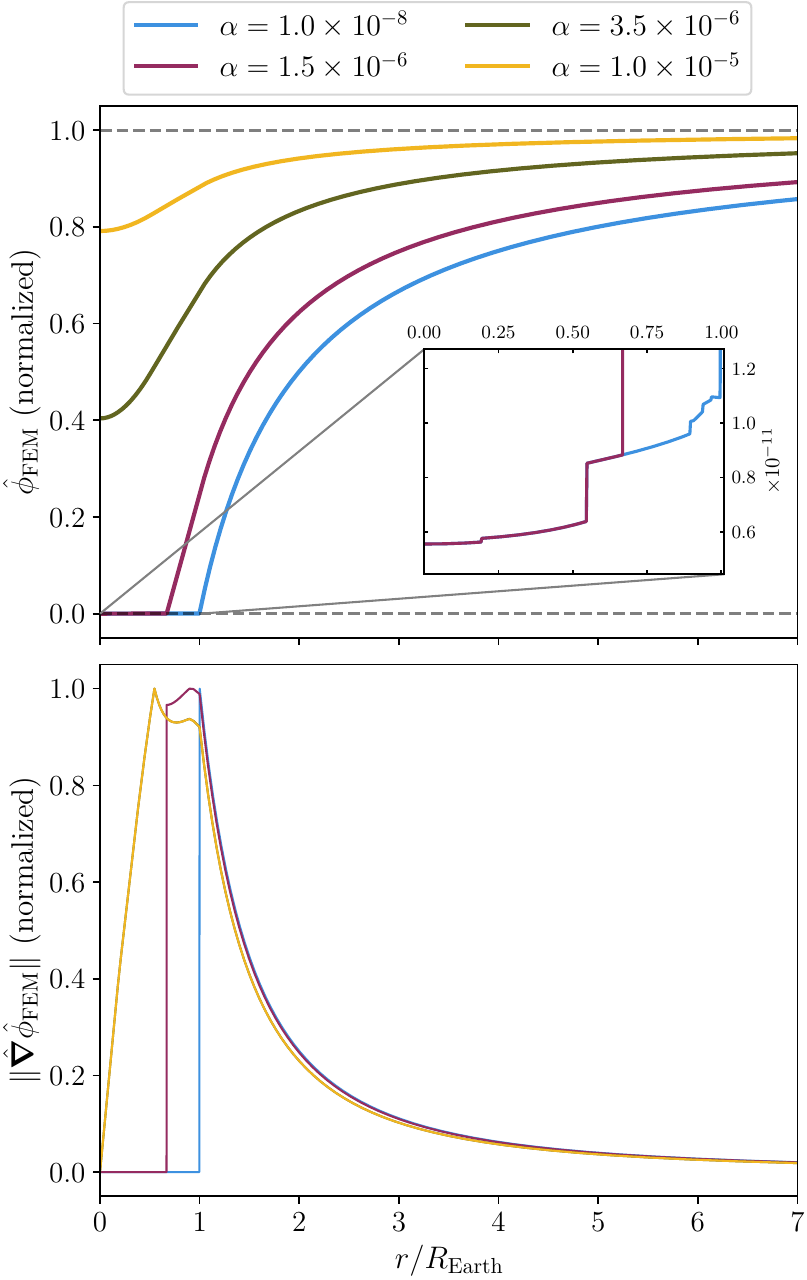}
    \caption{Chameleon field radial profile (top) and gradient (bottom) around a spherically symmetric Earth with density profiles depicted in Fig.~\ref{fig:dens_model_curves}.}
    \label{fig:1d_earth}
\end{figure}

In Fig.~\ref{fig:1d_earth}, we represent profiles of the chameleon field and its gradient for different values of the $\alpha$ parameter. The computed dimensionless field is further normalized in such a way that it tends to 1 at infinity, while the gradient is mapped onto $[0, 1]$ \textemdash \ which allows for a better comparison of the profiles. The $\alpha$-values are chosen so as to span over both the so-called screened regime ($\alpha \in \{ 10^{-8}, 1.5 \times 10^{-6} \}$) and unscreened regime ($\alpha \in \{ 3.5 \times 10^{-6}, 10^{-5} \}$). As can be seen on the inset, in the screened regime, the field is subject to jumps occurring at density jumps within the Earth, before stalling when the density crosses some threshold. This is the region where the corresponding gradient curve peaks to its highest value, before decreasing as $r^{-2}$ in the upper atmosphere and beyond. In the unscreened regime, the field does not reach the value that minimizes the effective potential at the center of the Earth. One point worth mentioning is that, in this regime, the fields' curves are identical up to an affine transformation. As a consequence, the associated normalized gradients almost perfectly overlap. A physical interpretation of this phenomenon is that in the unscreened regime, the field is sourced by the entire mass of the Earth and there is no thin-shell effect. The overall shape of the gradient is reminiscent of the Earth Newtonian gravity, which makes sense considering that the gravitational potential is not subject to any screening mechanism. Finally, let us denote by $\alpha_{\mathrm{screened}} \simeq 2.6 \times 10^{-6}$ the value at which the transition between the two regimes occurs\footnote{For $n=2$, one would have $\alpha_{\mathrm{screened}} \simeq 3.1 \times 10^{-3}$.}.

\begin{figure}
    \centering
    \includegraphics[width=0.65\textwidth]{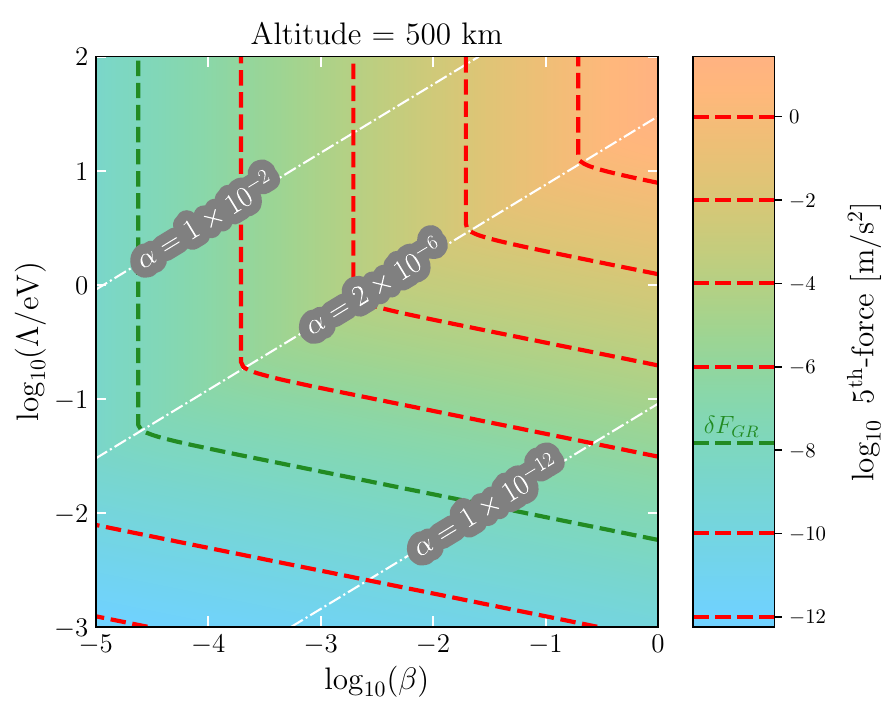}
    \caption{Analytical approximation of the $5^{\mathrm{th}}$-force value around an homogeneous, spherically symmetric Earth in the parameter-space $(\beta, \Lambda)$ for $n=1$. Iso-force values are depicted by dashed red lines and clearly exhibit two regimes delimited by the dimensionless parameter $\alpha \simeq 2 \times 10^{-6}$. The upper left region ($\alpha > 2 \times 10^{-6}$) corresponds to an unscreened Earth while the right region ($\alpha < 2 \times 10^{-6}$) corresponds to a screened Earth.}
    \label{fig:iso-f-ana}
\end{figure}

Let us now proceed to a more quantitative analysis of the chameleon field effects by computing the $\mathrm{5^{th}}$-force supposedly applied on satellite in orbit. This force is computed according to Eq.~(\ref{eqn:force-natural}). The mapping $(\beta , \Lambda) \mapsto \alpha$ not being injective, it is relevant to study the shape of the iso-$\mathrm{5^{th}}$-forces in the $(\beta, \Lambda)$-plane. Using the analytical approximation Eqs.~(\ref{eqn:analytical_screened}, \ref{eqn:analytical_unscreened}) is a good starting point to get a sketch of such contour lines. Because this approximation can only handle constant density profiles inside and outside the sphere, we separately average the PREM and US76 models depicted in Fig.~\ref{fig:dens_model_curves} and keep the two mean values. The result of this process is shown in Fig.~\ref{fig:iso-f-ana}, where we can clearly see the demarcation between the two regimes across the line $\alpha_{\mathrm{screened}} \simeq 2 \times 10^{-6}$. The Earth is screened (respectively unscreened) below (respectively above) this line. Note that we obtain the same characteristic iso-force contours as in Fig.~7 from Ref.~\cite{burrage_ellipticity} (plotted in the $(\log\Lambda, -\log\beta)$-plane).

It is striking to note that in the unscreened regime, the $\mathrm{5^{th}}$-force almost no longer depends on the energy scale $\Lambda$. This is particularly visible on the analytical approximation. From Eq.~(\ref{eqn:analytical_unscreened}), one has $\phi'(r) = K(1+m_{\mathrm{vac}}r) r^{-2} \exp[-m_{\mathrm{vac}}(r-R_A)]$. However, the vacuum density used in this study is so small ($\rho_{\mathrm{vac}} = 4.04 \times 10^{-19} \ \mathrm{kg/m^3}$) that, at a satellite altitude, $m_{\mathrm{vac}}r \ll 1$ and thus $\phi'(r) \sim K/r^2$, with $K \sim \rho_{\mathrm{in}}/3 \alpha$ so that $\phi'(r) \approxprop \Lambda^{(n+4)/(n+1)}$. The dimensionful version of the force is recovered by multiplying the dimensionless gradient by the factor $\beta \phi_0 / (\mathrm{M_{Pl}} L_0) \propto \Lambda^{(n+1)/(n+4)}$. Consequently, the result of this multiplication does not depend on $\Lambda$.
\begin{figure}
    \centering
    \includegraphics[width=0.6\textwidth]{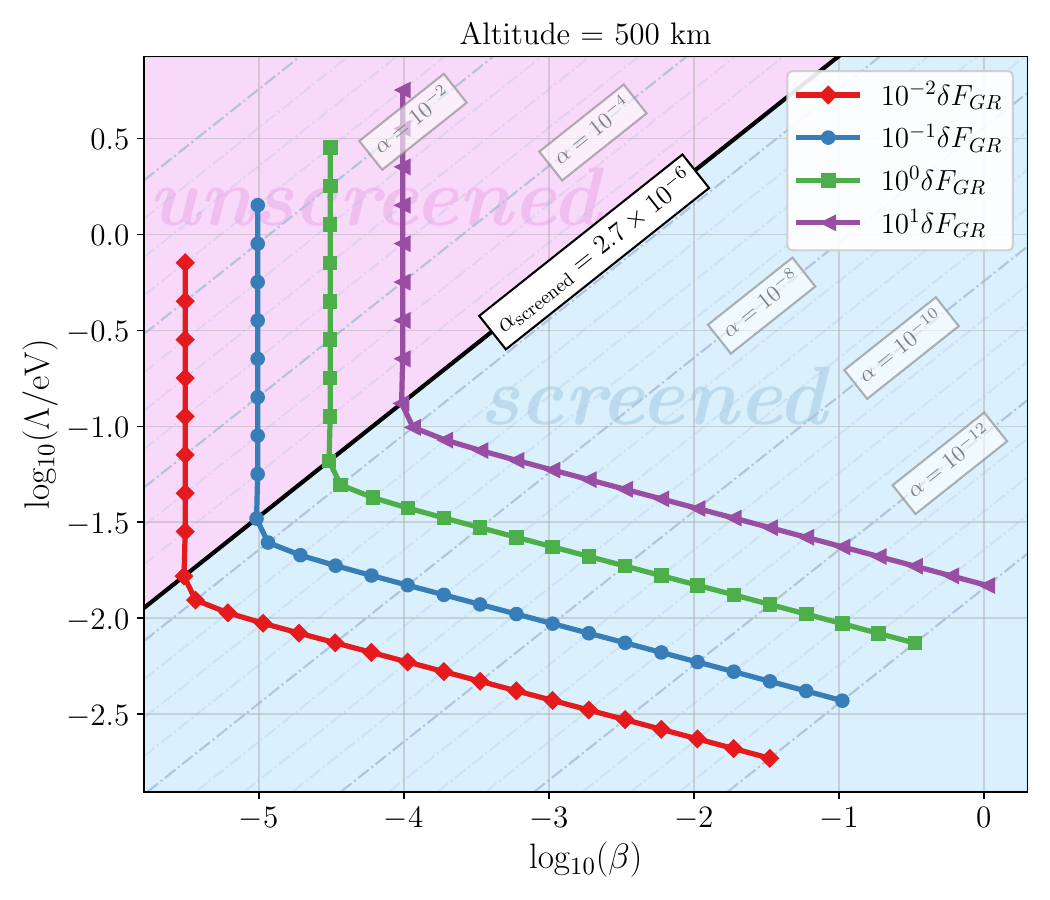}
    \caption{$5^{\mathrm{th}}$-force isolines computed on 1D FEM model with a realistic density model (see Fig. \ref{fig:dens_model_curves}). Iso-force lines are such that $F_{5^{\mathrm{th}}} = 10^{k} \delta F_{\mathrm{GR}}$, with $k$ ranging from -4 to 0 and $\delta F_{\mathrm{GR}}$ being the relativistic correction to the Newtonian equation of motion (see Ref.~\cite{satellite_orbits} for instance) given by Eq.~(\ref{eqn:delta_GR}). Gray lines in the background corresponds to the iso-values of the $\alpha$ dimensionless parameter used in the simulations. As one could foresee thanks to Fig. \ref{fig:iso-f-ana}, the screened regime (blue shade) and the unscreened regime (purple shade) are unmistakably separated on both sides of $\alpha_{\mathrm{screen}} \simeq 2.7 \times 10^{-6}$.}
    \label{fig:iso-f-num}
\end{figure}
The insights gained in the above helps us to comment on the results obtained with \emph{femtoscope} and the realistic density model. Fig.~\ref{fig:iso-f-num} is the numerical counterpart of Fig.~\ref{fig:iso-f-ana}, where we have represented the curves of equation $F_{\mathrm{cham}} = 10^k \delta F_{\mathrm{GR}}$ for $-2 \leq k \leq 1$. We obtain the same characteristic iso-$\mathrm{5^{th}}$-forces shape, whose equations roughly reads
\begin{equation}
    \begin{cases}
    \beta \sim \mathrm{const.} & \text{for } \alpha \geq \alpha_{\mathrm{screened}} \\
    \Lambda \sim \kappa \beta^{-1/5} & \text{for } \alpha < \alpha_{\mathrm{screened}}
    \end{cases}
\label{eqn:iso-f}
\end{equation}
for some positive constant $\kappa$. The power $-1/5$ can be recovered from the analytical approximation which gives $-n/(n+4)$ in the general case. This kind of plot has to be put into perspective with the current existing constraints on the chameleon model \cite{constraints, constraints_ahlers}. A space-based experiment in search of a chameleon $\mathrm{5^{th}}$-force at the level $10^k \delta F_{GR}$ could potentially rule out the whole upper right region of the corresponding curve on Fig.~\ref{fig:iso-f-num}. In Figs.~\ref{fig:iso-f-ana} and \ref{fig:iso-f-num}, we depict in green contour line along which the $\mathrm{5^{th}}$-force is equal to the relativistic correction given by Eq.~(\ref{eqn:delta_GR}).
\begin{figure*}
    \centering
    \includegraphics[width=0.9\textwidth]{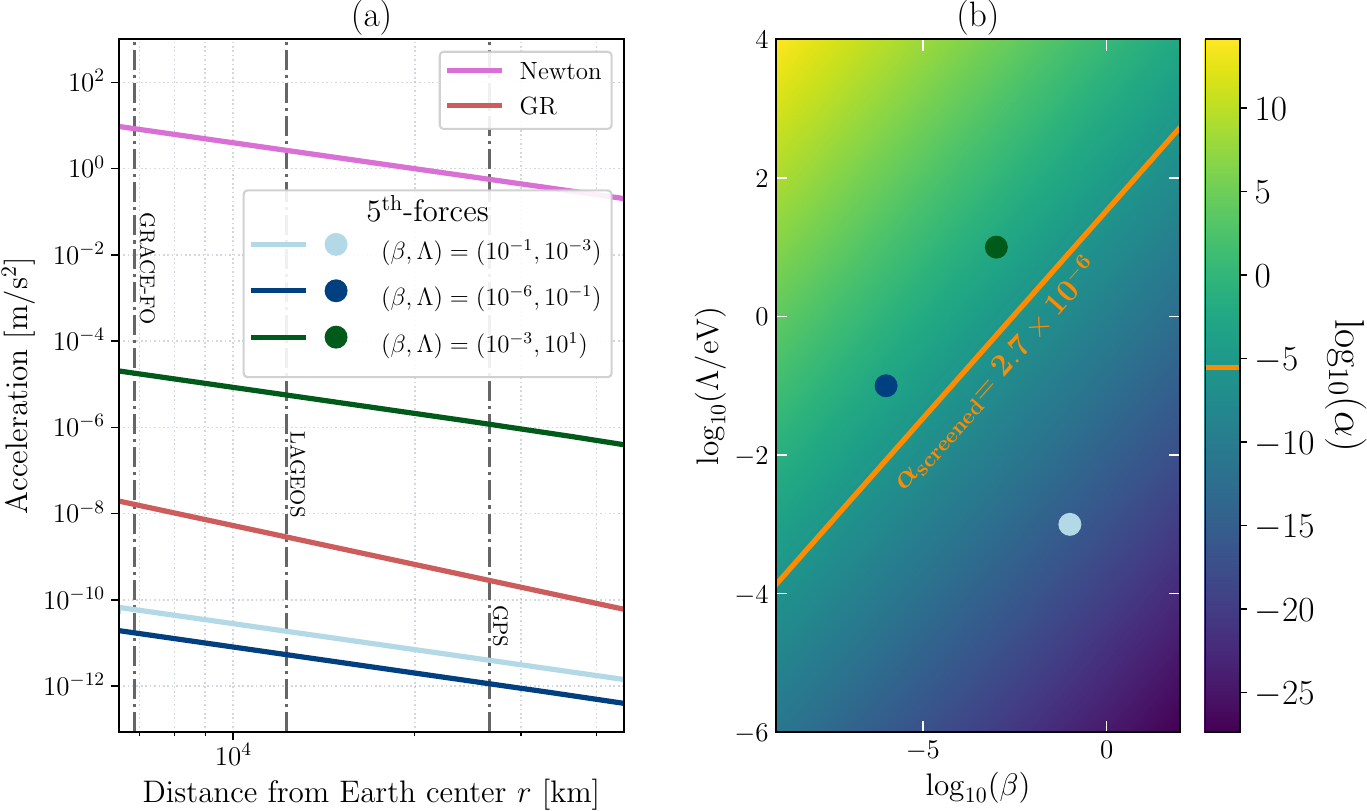}
    \caption{The chameleonic force as a perturbing acceleration for satellites. (a) Order of magnitude of hypothetical $\mathrm{5^{th}}$-forces alongside known forces (Newtonian gravity and its first order general relativistic correction) \textit{vs} $r$. The $(\beta , \Lambda)$ values used to compute the $\mathrm{5^{th}}$-forces are reported on the chameleon parameter space (b).}
\label{fig:f_vs_r}
\end{figure*}

Finally, it is useful to see how the chameleonic force compares to our current description of gravity in space. To that extent, we reproduced in Fig.~\ref{fig:f_vs_r} the traditional representation of satellite perturbations as a function of the altitude (see e.g. Fig.~3.1 from Ref.~\cite{satellite_orbits}). It features the Newtonian gravity $a_{\mathrm{Newton}} = \mu_{\mathrm{Earth}} / r^2$ and its relativistic correction at first order given by Eq.~(\ref{eqn:delta_GR}) as well as $\mathrm{5^{th}}$-force profiles. Yet, the two pairs $(\beta, \Lambda)$ which result in an unscreened Earth are already ruled-out by experiments \textemdash \ see, e.g., Fig.~2 from Ref.~\cite{constraints_ahlers}\footnote{To the best of our knowledge, there are no more up-to-date chameleon constraints' compilations in the range $\beta \in [10^{-9}, 10^{-3}]$.}. Overall, the region of the parameter space corresponding to an unscreened Earth is more severely constrained than the screened part. In the latter case, the freezing of the field inside the Earth means that the exterior field profile is sourced only by the mass outside the thin-shell radius. This puts into question the validity of a purely radial density model. Indeed, the shell sourcing the field might be so thin that it is no longer possible to make the assumption that the Earth is spherically-symmetric. In which case, it is reasonable to expect the $\mathrm{5^{th}}$-force to be dependent on the local landform. FEM would then be necessary to capture the aspherical shape of the topography.

A commentary has to be made with respect to the use of realistic physical quantities. Specifically, we noticed that numerical issues can arise when density varies widely within the simulation domain, like in Sec.~\ref{subsec:realistic-earth} where it ranges across 23 orders of magnitude. Part of the chameleon parameter space associated with an unscreened Earth ended up inaccessible to our numerical tool as the relative variation of the field $(\phi_{\mathrm{max}} - \phi_{\mathrm{min}}) / \phi_{\mathrm{max}}$ would be of order $\sim 10^{-14}$, close to machine epsilon in double-precision floating-point format.

\section{Conclusion \& Outlooks}
\label{sec:conclusion}

In this article, we have introduced \emph{femtoscope} as a novel numerical tool which is so far dedicated to the study of the chameleon model, a particular class of nonlinear scalar field theories. The interplay of self-interactions and coupling to matter makes for a rich phenomenology as the effective mass of the chameleon field becomes dependent on the local matter density. At the equation level, this translates into a nonlinear PDE that is difficult, if not impossible, to tackle by analytical means. Yet, both the design of an experiment and the analysis of the observations that come with it crucially hinge on our ability to make precision predictions of its outcome. This motivates the resort to numerical techniques in order to compute an approximation of the $\mathrm{5^{th}}$-force mediated by the chameleon field.

\emph{femtoscope} solves the chameleon field equation using the finite element method which is well-suited for our purposes as it can handle arbitrary matter distributions via non-uniform meshes. We implemented the traditional Newton's method (optionally with a line-search algorithm on top of it) in order to deal with the nonlinearity. The main novelty of this code lies in the implementation of various techniques to work on unbounded domains: one based on the compactification of the whole domain and two others based on domain splitting and Kelvin inversion. Overall, we have demonstrated the reliability of such techniques by using a special case of the Poisson equation (for which the exact solution is known) as a benchmark. In a way, our code extends the work by Briddon et al. \cite{Briddon_2021} by adding the possibility to deal with asymptotic boundary conditions. Furthermore, we underline the fact that \emph{femtoscope} is not inherently limited to the study of the chameleon field. Indeed, the numerical techniques discussed in this article rely only on a few assumptions, making \emph{femtoscope} a potential tool for dealing with a much wider range of PDEs. For instance, the symmetron model is another scalar-tensor model exhibiting yet a different screening mechanism that could well be studied with \emph{femtoscope}.

In this article, we conducted a study of the chameleon field and its underlying $\mathrm{5^{th}}$-force in the Earth environment. We started from the canonical case of a perfect homogeneous solid sphere (for which analytical approximations exist) before moving to a more realistic model of the matter distribution inside the Earth and in the atmosphere. This both showcases the possibilities offered by \emph{femtoscope} and paves the way to future, more case-specific, numerical studies of the chameleonic force. Specifically, we have underlined the fact that modeling the Earth as a sphere is no longer realistic in the screened regime where the chameleonic force is sourced by the outer layers. It would be interesting to see the imprint of the local relief on the chameleon field in Earth orbits and whether or not we could discriminate between the $\mathrm{5^{th}}$-force signature and known effects with the current technology embedded on navigation and potential science satellites. Finally, many astrophysical systems boil down to the study of a two-body problem. The nonlinearity of the chameleon field equation prevents from considering each body independently from the other and applying the principle of superposition. Instead, a full 3D computation of the field is required due to the absence of symmetry of such a setup. Moreover, having two bodies \textemdash \ say a binary neutron stars \textemdash \ orbiting each other breaks the assumption of a static scalar-field and one would have to solve the full, time-dependent, Klein-Gordon equation. This feature is not yet implemented in \emph{femtoscope} and is left for future work.

\begin{acknowledgments}
We thank Dr. Robert Cimrman (University of West Bohemia, Czech Republic) for providing assistance with the use of \textit{Sfepy}.
\end{acknowledgments}

\appendix

\section{Comparison between SELCIE and \emph{femtoscope}}
\label{sec:selcie_vs_femtoscope}

\begin{figure}
    \centering
    \includegraphics[width=0.8\textwidth]{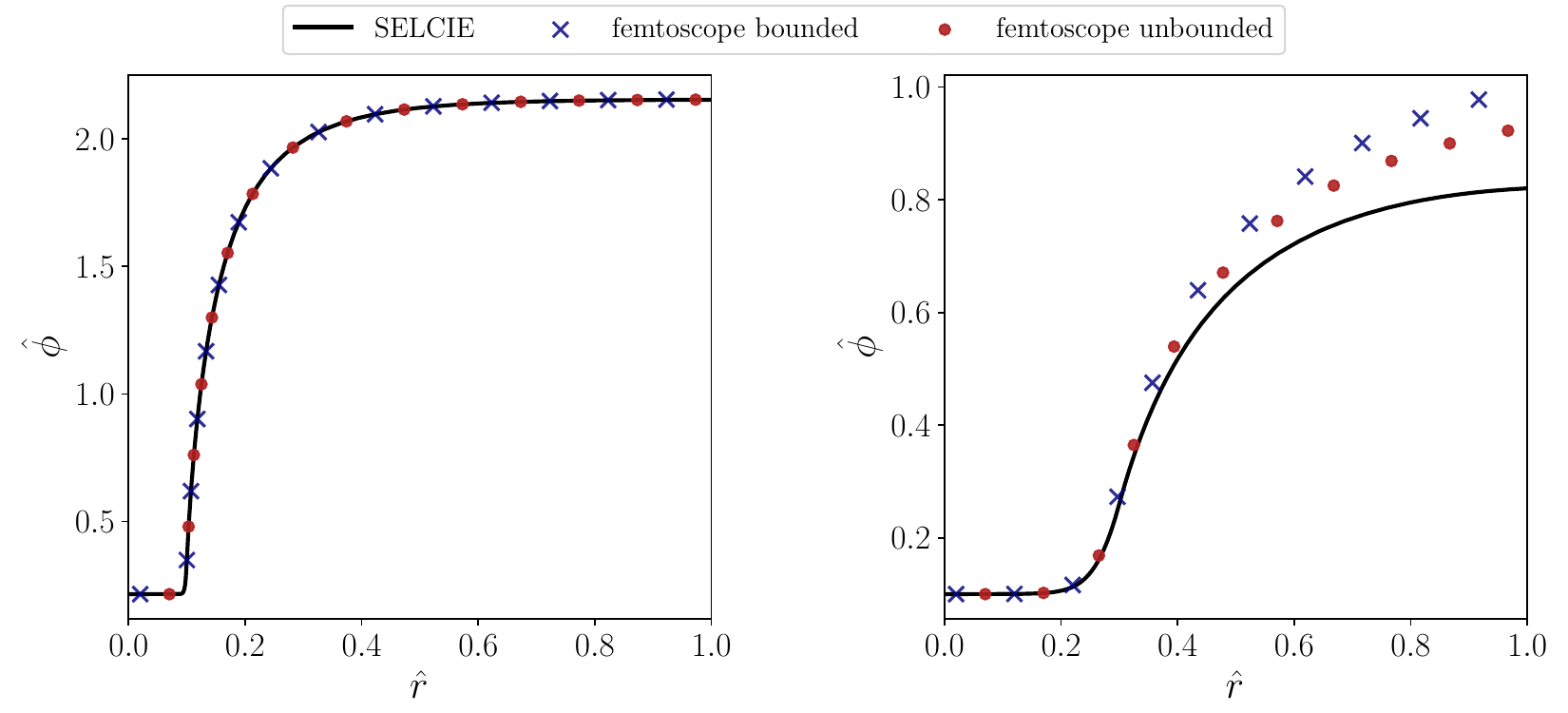}
    \caption{Comparison between SELCIE (black solid line) and \emph{femtoscope}, (i) with Dirichlet boundary condition at the truncation radius $R_{\mathrm{cut}} = 1$ (blue crosses) and (ii) with the virtual connection of DOFs technique (red dots). The parameters used are: \{$n=2$, $\alpha=5 \times 10^{-3}$, $R_{\mathrm{cut}} = 1$, $R_{A} = 0.1$, $\rho_{\mathrm{in}} = 100$, $\rho_{\mathrm{vac}} = 0.1$\} for the left panel and \{$n=1$, $\alpha=1$, $R_{\mathrm{cut}} = 1$, $R_{A} = 0.3$, $\rho_{\mathrm{in}} = 100$, $\rho_{\mathrm{vac}} = 1$\} for the right  one. The SELCIE simulations were performed by adapting an existing example provided on its GitHub repository\footnote{\url{https://github.com/C-Briddon/SELCIE/blob/main/Examples/Solve_SphereCylinder.py}. Last visited: August $\mathrm{1^{st}}$, 2022.}.}
    \label{fig:selcie_vs_femtoscope}
\end{figure}

To the best of our knowledge, SELCIE \cite{Briddon_2021} is the only publicly available code that can be used to compute the chameleon field for arbitrary density distributions. Despite sharing many similarities with SELCIE, \emph{femtoscope} was developed in an independent way to achieve close aims. It is therefore all the more important to check that the two codes' output coincide on a given simulation as no analytical solution is available. We selected a set of physical parameters and computed the chameleon field for a perfect solid sphere with the two softwares. Unlike \emph{femtoscope}, SELCIE is not able to deal with asymptotic boundary conditions. Instead by default\footnote{The initial field profile can also be user supplied since version 1.4.0.}, it initializes the field to $\hat{\phi}_{\mathrm{min}}$ which is computed from the maximum density within the domain $\hat{\rho}_{\mathrm{max}}$ via the relation
\begin{equation}
    \hat{\phi}_{\mathrm{min}} = \hat{\rho}_{\mathrm{max}}^{-\frac{1}{n+1}} \, .
\label{eqn:phi_min}
\end{equation}
The artificial border at the truncation radius is left free of any Dirichlet boundary condition, hence natural boundary condition applies:
\begin{equation}
    \hat{\boldsymbol{\nabla}} \hat{\phi} \cdot \mathbf{n} = 0 \, ,
\label{eqn:neumann}
\end{equation} 
which has no physical reality in the general case. Fig.~\ref{fig:selcie_vs_femtoscope} aims at comparing SELCIE and \textit{femtoscope} on two different simulations performed in 2D Cartesian coordinates. Because \emph{femtoscope} is not limited to bounded domains, we solved the Klein-Gordon equation by way of two techniques:
\begin{enumerate}
    \item applying the Dirichlet boundary condition $\hat{\phi} = \hat{\phi}_{\mathrm{vac}}$ at the artificial border (blue crosses), or;
    \item using the virtual connection of DOFs described in Sec.~\ref{subsubsec:kelvin} to enforce the correct asymptotic behavior of the field at infinity (red dots). We recall once more that this is the most general method as no particular assumptions have to be made regarding the physical parameters of the simulation.
\end{enumerate}

\textbf{Left panel.} In order for SELCIE to produce a reasonable numerical approximation, we chose a set of physical parameters such that:
\begin{itemize}
    \item[--] the ball is screened. In this manner, the field is correctly initialized deep inside the ball via Eq.~(\ref{eqn:phi_min});
    \item[--] the field value at the truncation boundary is close to value that minimizes the effective potential outside the ball, denoted $\hat{\phi}_{\mathrm{vac}}$. As a result, the field's gradient is expected to be small and Eq.~(\ref{eqn:neumann}) makes physical sense.
\end{itemize}
The three outputs \textemdash \ SELCIE, \emph{femtoscope} bounded and unbounded \textemdash \ almost perfectly overlap. Indeed, the relative difference between any two of the three numerical approximations is bounded below 0.3\%. There are several potential causes to explain this sub-percentage difference:
\begin{itemize}
    \item[--] We did not use the same meshes for SELCIE and \emph{femtoscope}. However, the quality of the FEM solution is known to be intimately interrelated with that of the associated mesh \cite{mesh_quality}.
    \item[--] For all three outputs, the field is initialized and, more importantly, constrained differently (see the discussion above).
    \item[--] SELCIE and \emph{femtoscope} do not use the same linear solver on these specific simulations.
    \item[--] The Finite Element approximation order is set to one in SELCIE (by default) whereas we used third-order polynomials for the computations performed with \emph{femtoscope}.
\end{itemize}

\textbf{Right panel.} This plot aims at showing the limits of domain truncation. Here, \textit{femtoscope} unbounded  (red dots) should be regarded as the benchmark as it is the only simulation that correctly implements the asymptotic boundary condition. We can see that, as we move away from the ball ($\hat{r} \geq 0.3$), the three outputs start diverging:
\begin{itemize}
    \item[--] SELCIE (black solid line) implements condition (\ref{eqn:neumann}) which is why we observe $\left. \frac{\mathrm{d} \hat{\phi}}{\mathrm{d}\hat{r}}\right|_{\hat{r}=1} = 0$. This is wrong because the field should keep increasing to $\hat{\phi}_{\mathrm{vac}}$ at infinity and results in a significant deviation from the benchmark (relative difference up to 11\% at $\hat{r}=1$).
    \item[--] \textit{femtoscope} bounded (blue crosses) implements the Dirichlet boundary condition $\hat{\phi}(\hat{r}=1) = \hat{\phi}_{\mathrm{vac}}$. This results in a 6\% relative difference at $\hat{r}=1$.
\end{itemize}

\begin{figure}[t]
    \centering
    \includegraphics[width=0.45\textwidth]{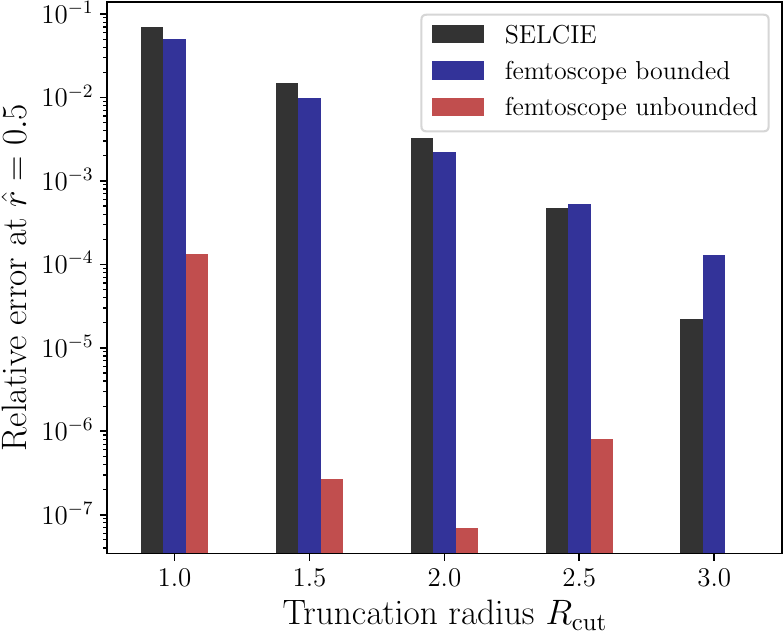}
    \caption{Influence of the truncation radius on the error. The simulations were performed with the set of parameters \{$n=1$, $\alpha=1$, $R_{A} = 0.3$, $\rho_{\mathrm{in}} = 100$, $\rho_{\mathrm{vac}} = 1$\} (same as the right panel of Fig.~\ref{fig:selcie_vs_femtoscope}). Here, the chosen benchmark is \{\textit{femtoscope} unbounded, $R_{\mathrm{cut}}=3.0$\}.}
    \label{fig:truncation-study}
\end{figure}

Finally, we complement this comparative study by addressing the question of the influence of the size of the truncated domain on accuracy. To do so, we start from $R_{\mathrm{cut}} = 3$ and compute $\hat{\phi}(\hat{r}=0.5)$ using the virtual connection of DOFs technique: this is our benchmark. Then we decrease the truncation radius down to 1 in steps of 0.5 and compute the relative error at $\hat{r}=0.5$ for all three outputs. The results of this experiment are shown in Fig.~\ref{fig:truncation-study}. The takeaway here is that truncation techniques (SELCIE and \textit{femtoscope} bounded) become increasingly inaccurate as $R_{\mathrm{cut}}$ decreases. Moreover, for an arbitrary set of parameters, the truncation radius ensuring an acceptable level of error cannot in principle be known. One thus has to be very cautious when using codes relying on truncation and have enough physical insights into how to choose the truncation radius. As for \textit{femtoscope} unbounded, the dependence between the error and $R_{\mathrm{cut}}$ is much less pronounced, except for $R_{\mathrm{cut}} = 1$ where the relative error goes beyond $10^{-4}$. This brief investigation, although it is merely one example, further illustrates why properly dealing with boundary conditions is of key importance.

\section{Relation between $\alpha$ and the residual size}
\label{sec:alpha-res_relation}

\begin{figure}[H]
    \centering
    \includegraphics[width=0.78\textwidth]{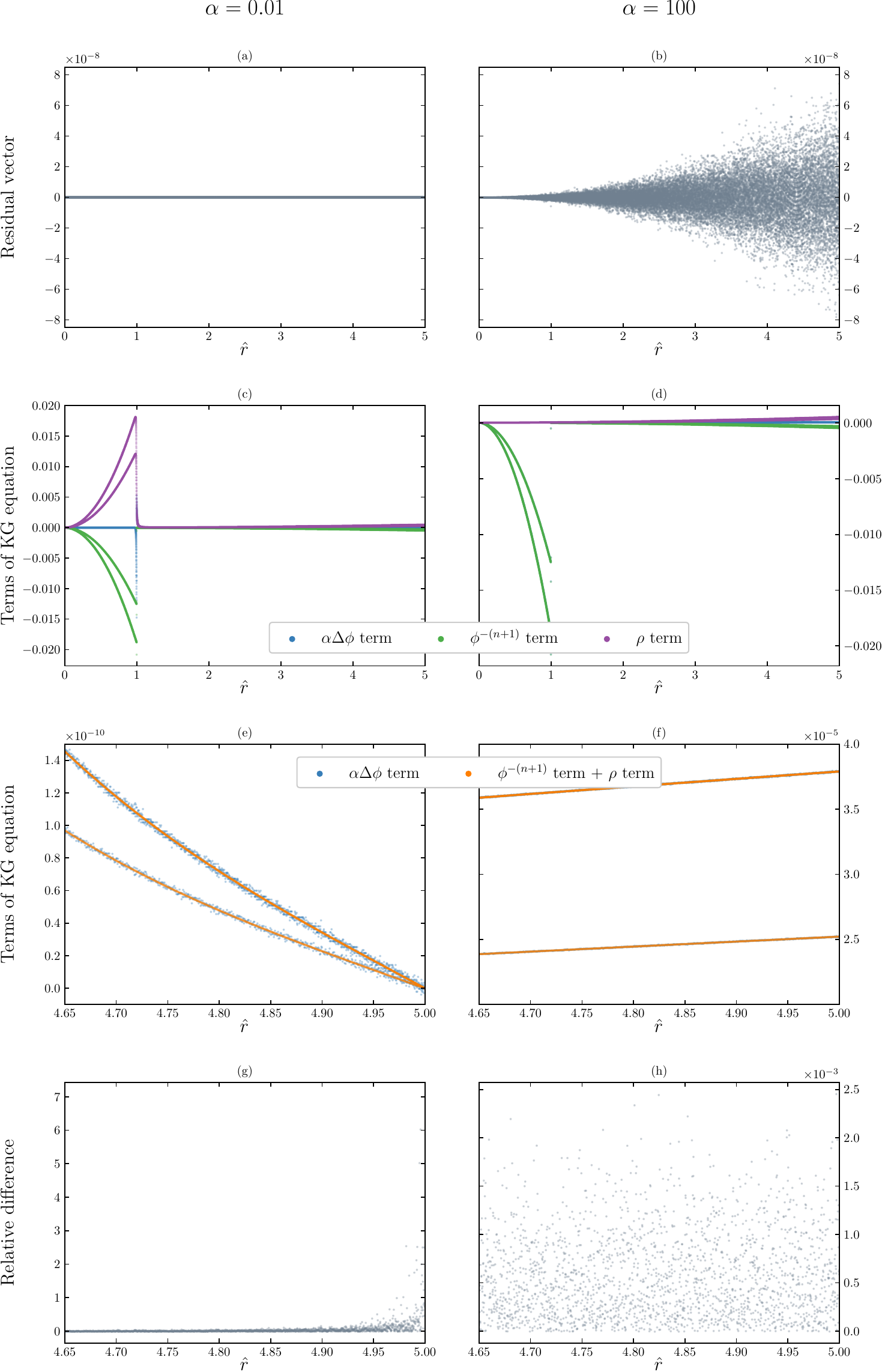}
    \caption{Investigation of the relationship between the $\alpha$ parameter and the residual of the numerical approximation after convergence has been reached (25 iterations). Left column: $\alpha = 0.01$. Right column: $\alpha= 100$. See the main text for a more detailed description of each panel.}
    \label{fig:alpha-res}
\end{figure}

This appendix follows on from the discussion initiated in Sec.~\ref{subsec:verif-case} and proposes an explanation for the apparent relationship between the $\alpha$ parameter and the residual (see Table~\ref{tab:res-ana-num}). We investigated the residual vector for $\alpha \in \{ 10^{-2}, 10^{2} \}$ on 1D simulations and the important results are shown in Fig.~\ref{fig:alpha-res}. The left-column plots correspond to $\alpha = 0.01$ while the right-column plots correspond to $\alpha = 100$. We have indeed deliberately chosen two values of $\alpha$ separated by several orders of magnitude so as to accentuate the trend observed in Table~\ref{tab:res-ana-num}. The first line of Fig.~\ref{fig:alpha-res} (panels a and b) depicts the residual vector as a function of the dimensionless radial coordinate $\hat{r}$ (the two plots have the same y-scale). For both values of $\alpha$, the residual vector is almost consistent with 0 (the mean residual is $-3 \times 10^{-15}$ for $\alpha = 0.01$ and $-8 \times 10^{-11}$ for $\alpha = 100$), but the dispersion is much greater for $\alpha = 100$ (the standard deviation is more than four orders of magnitude bigger). Two questions arise:
\begin{enumerate}
    \item What is this huge dispersion due to?
    \item Why does the dispersion increase with $\hat{r}$?
\end{enumerate}
Schematically, the residual vector is evaluated by plugging the numerical approximation $\phi^h$ into the Klein-Gordon equation (\ref{eqn:kg3}) and see how small the quantity $\alpha \Delta \phi^h - [-\left(\phi^h\right)^{-(n+1)}] - \rho $ is (a proper definition of the residual vector is given by Eq.~\ref{eqn:discrete-nonlinear}). To better understand the influence of $\alpha$ on the residual, it is insightful to decompose the residual into three terms \textemdash \ the $\alpha \Delta \phi$ term, the $\phi^{-(n+1)}$ term and the density, or $\rho$, term \textemdash \ and plot each of them against $\hat{r}$. This is done on panels (c) and (d). We then compare the $\alpha$-dependent part against the two other terms (panels e and f) and compute their relative difference difference with respect to the $\alpha \Delta \phi$ term (panels g and h). We observe that:
\begin{itemize}
    \item[--] The relative difference between the $\alpha \Delta \phi$ term and $(\rho - \phi^{-(n+1)})$ is smaller for $\alpha=100$ than for $\alpha=0.01$ (panels g and h).
    \item[--] Yet, in absolute terms, these two terms are about five orders of magnitude bigger for $\alpha = 100$ than for $\alpha = 0.01$ (panels e and f). This is most likely due to the fact that $\alpha$ appears in the Klein-Gordon equation as a multiplicative constant before the Laplacian.
\end{itemize}
Consequently, the difference between these two terms (which is nothing but the residual) is several orders of magnitude larger (panels a and b) as expected. This answers question 1. The second question is more straightforward to answer. Recall that in 1D, the Laplacian operator is expressed as $$ \Delta \phi = \frac{1}{r^2} \frac{\mathrm{d}}{\mathrm{d}r} \left(r^2 \frac{\mathrm{d} \phi}{\mathrm{d}r} \right) \, ,$$ and we multiply both sides of the Klein-Gordon equation by $r^2$ before deriving the weak formulation, so that the $\rho$ and $\phi^{-(n+1)}$ are weighted by $r^2$. As a result, the terms involved in the residual get larger in absolute value as $\hat{r}$ increases, which is why the dispersion of the residual gets wider as we go to larger $\hat{r}$ values (plot b).

\section{Mathematical proof of the vanishing gradient}
\label{sec:bc-proof}

Here we provide a mathematical proof that 
$$ \begin{cases}
  \alpha \Delta \phi = \rho - \phi^{-(n+1)}\\[5pt]
  \phi \underset{\|\mathbf{x}\| \to +\infty}{\longrightarrow} \phi_{\mathrm{vac}}\\[15pt]
  \partial_{\theta} \phi$, $\partial_{\varphi} \phi$, $\partial_{\theta}^2 \phi$, $\partial_{\varphi}^2 \phi = O_{\|\mathbf{x}\| \to \infty}(1)
\end{cases}  \implies \| \boldsymbol{\nabla} \phi \| \underset{\|\mathbf{x}\| \to +\infty}{\longrightarrow} 0 \, .$$
The asymptotic value of the field is known to be $\phi_{\mathrm{vac}} = (\rho_{\mathrm{vac}})^{-\frac{1}{n+1}}$ so that the r.h.s. of Eq.~(\ref{eqn:model_pb}) goes to zero at infinity, i.e.
$$ \Delta \phi(\mathbf{x}) \underset{\|\mathbf{x}\| \to +\infty}{\longrightarrow} 0 \, . $$
In spherical coordinates, the Laplacian of $\phi : \mathbb{R}^3 \to \mathbb{R}$ reads
$$ \Delta \phi \coloneqq \underbrace{\frac{1}{r^2} \frac{\partial}{\partial r} \left( r^2 \frac{\partial \phi}{\partial r} \right)}_{\text{radial part}} + \underbrace{\frac{1}{r^2 \sin (\theta)} \frac{\partial}{\partial \theta} \left( \sin (\theta) \frac{\partial \phi}{\partial \theta} \right) + \frac{1}{r^2 \sin^2(\theta)} \frac{\partial^2 \phi}{\partial \varphi^2}}_{\text{angular part}} \, .$$
Because of the assumptions we have made on the partial derivatives involving the angles $(\theta, \varphi)$, the angular part of the Laplacian and gradient vanish to zero as $r$ goes to infinity so that we can focus our attention on the radial part. In order to simplify the notation, from now on we consider $\phi$ to be purely radial and we denote by $\phi'$ and $\phi''$ its first and second order derivatives with respect to $r$ respectively. The vanishing Laplacian reduces to
\begin{equation}
    \frac{1}{r^2} \frac{\mathrm{d}}{\mathrm{d}r} \left( r^2 \frac{\mathrm{d}\phi}{\mathrm{d}r} \right) = \phi''(r) + \frac{2}{r} \phi'(r) \underset{r \to +\infty}{\longrightarrow} 0 \, .
\label{eqn:radial-laplacian}
\end{equation}

\subsection{Proof that \texorpdfstring{$\mathbf{\phi''(r) \underset{r \to +\infty}{\longrightarrow} 0}$}{the second order derivative goes to zero}}

The asymptotic condition on the radial part of the Laplacian (\ref{eqn:radial-laplacian}) may be reformulated as:
$$ \text{there exists a function} \ \epsilon : \mathbb{R}_+^* \to \mathbb{R} \ \text{such that} \ \begin{cases}
\phi''(r) + \dfrac{2}{r} \phi'(r) = \epsilon(r) \\[10pt]
\epsilon(r) \underset{r \to +\infty}{\longrightarrow} 0
\end{cases} \, . $$
The above is nothing but a second-order linear ordinary differential equation (ODE) which can be solved via the method of variation of parameters. The general solution of the homogeneous equation can be expressed as $-A/r + B$, with $A, B \in \mathbb{R}$. Then a particular solution of the inhomogeneous equation is sought in the form $\phi(r) = -A(r)/r + B(r)$, with $A$ and $B$ two real functions satisfying the system
$$ \begin{cases}
-A'(r)/r + B'(r) = 0 \\
A'(r)/r^2 + 0 = \epsilon(r)
\end{cases} \iff
\begin{cases}
A'(r) = r^2 \epsilon(r) \\
B'(r) = r \epsilon(r)
\end{cases} \, .$$
Therefore, a particular solution of the ODE on $\mathbb{R}_+^*$ is
$$ \phi(r) = -\frac{1}{r} \int_{1}^{r} s^2 \epsilon(s) \, \mathrm{d}s + \int_1^r s \epsilon(s) \, \mathrm{d}s \, . $$
The general solution then reads
$$ \phi(r) = -\frac{1}{r} \left[ \int_{1}^{r} s^2 \epsilon(s) \, \mathrm{d}s + A \right] + \int_1^r s \epsilon(s) \, \mathrm{d}s + B \quad , \quad A, B \in \mathbb{R} \, . $$
From there, we can compute the second order derivative as
$$ \phi''(r) = -\frac{2}{r^3} \left[ \int_1^r s^2 \epsilon(s) \, \mathrm{d}s + A \right] + \epsilon(r) \ $$
and the proof boils down to showing that
$$ \frac{1}{r^3} \int_1^r s^2 \epsilon(s) \, \mathrm{d}s \underset{r \to +\infty}{\longrightarrow} 0 \, . $$
Let $\delta > 0$, $\epsilon(r) \underset{r \to +\infty}{\longrightarrow} 0$ hence there exists $R_{\delta} > 0$ such that for all $r \geq R_{\delta}$, $|\epsilon(r)| < \delta$. Let us introduce
$$M \coloneqq \max_{s \in [1, +\infty[} |\epsilon(s)| \quad \text{and} \quad R_* \coloneqq \frac{R_{\delta} M}{\delta} \, .$$
For $r \geq \max(R_*, R_{\delta}) \eqqcolon R_{\mathrm{m}}$, we get:
\begin{flalign*}
\left|I(r)\right| \coloneqq \Bigg| \frac{1}{r^3} \int_1^r s^2 \epsilon(s) \, \mathrm{d}s \Bigg| &= \Bigg| \frac{1}{r} \int_1^r \underbrace{\left( \frac{s}{r} \right)^2}_{\leq 1} \epsilon(s) \, \mathrm{d}s \Bigg| \leq \frac{1}{r} \int_1^r |\epsilon(s)| \, \mathrm{d}s \\[10pt]
& \leq \frac{1}{r} \int_1^{R_{\delta}} |\epsilon(s)| \, \mathrm{d}s + \frac{1}{r} \int_{R_{\delta}}^r |\epsilon(s)| \, \mathrm{d}s \\[10pt]
& \leq \frac{1}{r} \int_1^{R_{\delta}} \max_{s \in [1, R_{\delta}]} |\epsilon(s)| \, \mathrm{d}s + \frac{1}{r} \int_{R_{\delta}}^r \delta \, \mathrm{d}s \\[10pt]
& \leq \frac{R_{\delta}-1}{r} \max_{s \in [1, R_{\delta}]} |\epsilon(s)| + \frac{r-R_{\delta}}{r} \delta \\[10pt]
& \leq \frac{R_{\delta}M}{r} + \delta \leq \frac{R_{\delta}M}{R_*} + \delta \leq \delta + \delta \leq 2 \delta \, .
\end{flalign*}
We have shown that $\forall \delta > 0 \ , \ \exists R_{\mathrm{m}}>0 \ / \ \forall r>R_{\mathrm{m}} \ , \left| I(r) \right| \leq \delta$, which is the exact definition of $I(r) \underset{r \to +\infty}{\longrightarrow} 0$ and concludes the first part of the proof.

\subsection{Proof that \texorpdfstring{$\mathbf{\phi'(r) \underset{r \to +\infty}{\longrightarrow} 0}$}{the first order derivative goes to zero}}

Let $f \in \mathcal{C}^2(\mathbb{R}_+, \mathbb{R})$ be such that
\begin{equation}
   \begin{cases}
f \ \text{has a limit} \ l \ \text{as } x \text{ approaches} \ +\infty \\
f'' \ \text{goes to 0 as } x \text{ approaches} \ +\infty
\end{cases} \, .
\label{eqn:hyp}
\end{equation}
These two hypotheses can be rewritten in a more mathematical formalism as
\begin{equation}
    \text{[} f'' \text{ goes to 0]} \quad \forall \epsilon > 0 \ , \ \exists M \in \mathbb{R}_+ \ / \ \forall x \geq M \ , \ |f''(x)| \leq \epsilon \, ,
\label{eqn:2nd_to_0}
\end{equation}
\begin{equation}
    \text{[} f \text{ goes to } l \text{]} \quad \forall \epsilon > 0 \ , \ \exists M \in \mathbb{R}_+ \ / \ \forall x \geq M \ , \ |f(x)-l| \leq \epsilon \, .
\label{eqn:f_to_l}
\end{equation}
The fact that $f$ converges allows us to write a third proposition that slightly differs from (\ref{eqn:f_to_l})
\begin{equation}
    \text{[} f \text{ converges]} \quad \forall \epsilon > 0 \ , \ \exists M \in \mathbb{R}_+ \ / \ \forall x_1, x_2 \geq M \ , \ |f(x_1)-f(x_2)| \leq \epsilon \, .
\label{eqn:f_conv}
\end{equation}
\textit{Strategy:} We develop a proof by contradiction. To that end, let us suppose that $f'$ does not go to 0 at $+ \infty$, that is
\begin{equation}
    \exists \delta > 0 \ / \ \forall A \in \mathbb{R}_+ \ , \ \exists x \geq A \ / \ |f'(x)| > \delta \, .
\label{eqn:noconv}
\end{equation}
Property (\ref{eqn:noconv}) provides us with $\delta > 0$. Even if it means redefining $f \gets -f$, one can get rid of the absolute value in (\ref{eqn:noconv}) so that
\begin{equation}
   \forall A \in \mathbb{R}_+ \ , \ \exists x \geq A \ / \ f'(x) > \delta \, .
\label{eqn:noconvbis}
\end{equation}
Note that this potential change of sign does not change in any way the asymptotic behavior of $f'$ and $f''$. From here, the proof follows the subsequent steps.
\begin{enumerate}
    \item $f'$ reaches $\delta$ for arbitrarily large $x$.
\end{enumerate}
More precisely, let us demonstrate that $\forall A > 0 \ , \ \exists x \geq A \ / \ f'(x) = \delta$. Let $A > 0$, according to (\ref{eqn:noconvbis}), there exists $x_m \geq A$ such that $f'(x_m) > \delta$. We employ reductio ad absurdum, assuming that for all $x \geq x_m$, $f'(x) \neq \delta$. Because $f'$ is continuous over $\mathbb{R}_+$, this implies that $\forall x \geq x_m$, $f'(x) > \delta$. This statement is in contradiction with the convergence of $f$. Indeed, let $\epsilon > 0$ and get $M \in \mathbb{R}_+$ given by property (\ref{eqn:f_conv}). We set
$$ x_1 \coloneqq \max(x_m, M) \quad \text{and} \quad x_2 \coloneqq x_1 + \frac{2}{\delta} \epsilon \, .$$
On the one hand,
$$ |f(x_1) - f(x_2)| \leq \epsilon \quad \text{because } x_1, x_2 \geq M \, , $$
and on the other hand, $\forall x \in [x_1, x_2] , \ f'(x) \geq \delta$ so that the mean value inequality gives
$$ \int_{x_1}^{x_2} f'(x) \, \mathrm{d}x \geq \int_{x_1}^{x_2} \delta \, \mathrm{d}x \quad \text{thus} \quad |f(x_1) - f(x_2)| \geq f(x_2)-f(x_1) \geq \delta |x_2 - x_1| = \delta \frac{2}{\delta} \epsilon = 2\epsilon >0 \, .$$
The contradiction is now clear.
\begin{enumerate}
\setcounter{enumi}{1}
    \item $f'$ reaches $\delta/2$ for arbitrarily large $x$.
\end{enumerate}
Using the exact sames arguments as above, one proves that  $\forall A > 0 \ , \ \exists x \geq A \ / \ f'(x) = \delta/2$. Before going any further, we define two sets:
$$ E_{\delta} \coloneqq \left\{ x \in \mathbb{R}_+ \ \text{such that} \ f'(x) = \delta \right\} \quad \text{and} \quad E_{\delta/2} \coloneqq \left\{ x \in \mathbb{R}_+ \ \text{such that} \ f'(x) = \frac{\delta}{2} \right\} \, . $$
We have just shown that these two sets are infinite and that they contain arbitrarily large values of $x$.
\begin{enumerate}
\setcounter{enumi}{2}
    \item Construction of the interval sequence  $(I_n)_{n \in \mathbb{N}}$.
\end{enumerate}
The aim of this part is to show that $f'$-values stay between $\delta/2$ and $\delta$ on arbitrarily large intervals. To that extent, we construct a sequence of disjoint intervals $(I_n)_{n \in \mathbb{N}}$ such that $f'$ falls between $\delta/2$ and $\delta$ on each $I_n$:
\begin{itemize}
    \item[--] For $I_0$, we set $x_{0, \delta}$ in $E_{\delta}$ and $x_{0, \delta/2}$ in $E_{\delta/2}$ such that $x_{0, \delta} < x_{0, \delta/2}$ and $\forall x \in [x_{0, \delta}, x_{0, \delta/2}], \ f'(x) \in [\delta/2 , \delta]$.
    
    \item[--] For $I_1$, we choose $x_{1, \delta}$ in $E_{\delta} \cap ]x_{0, \delta} +\infty]$ and $x_{1, \delta/2}$ in $E_{\delta/2} \cap ]x_{0, \delta/2} +\infty]$ such that $x_{1, \delta} < x_{1, \delta/2}$ and $\forall x \in [x_{1, \delta}, x_{1, \delta/2}], \ f'(x) \in [\delta/2 , \delta]$. By construction, $I_1$ and $I_0$ are indeed disjoints.
    
    \item[--] For $I_2$, we choose $x_{2, \delta}$ in $E_{\delta} \cap ]x_{1, \delta} +\infty]$ and $x_{2, \delta/2}$ in $E_{\delta/2} \cap ]x_{1, \delta/2} +\infty]$ such that ...
    
    \item[--] etc.
\end{itemize}
\begin{figure}[b]
    \centering
    \includegraphics[width=0.5\textwidth]{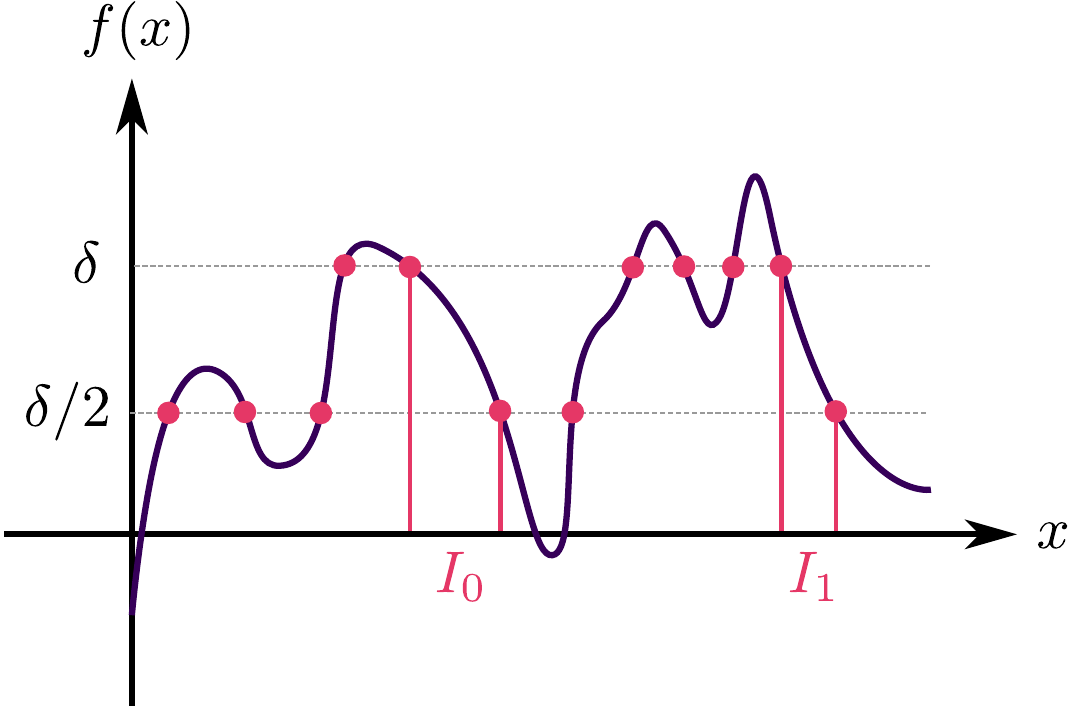}
    \caption{Construction of the $(I_n)_{n \in \mathbb{N}}$ sequence.}
    \label{fig:sequence}
\end{figure}
This construction is illustrated on Fig. \ref{fig:sequence}. We now demonstrate that
$$ \forall X, A > 0, \exists I \in (I_n)_{n \in \mathbb{N}} \ \text{such that} \ \begin{cases}
\inf (I) \geq X \\
|I| \geq A
\end{cases} \, . $$
Let $X, A > 0$ and set $\epsilon = A^{-1}$. We make use of the fact that $f''$ goes to 0 by applying property (\ref{eqn:2nd_to_0}) for $\epsilon \delta / 2 > 0$. Let us denote $M \geq 0$ the constant provided with this property and set $R \coloneqq \max (X, M)$. According to what has been shown in the previous point, one can choose an element $I = [a, b]$ of the sequence $(I_n)_{n \in \mathbb{N}}$ such that $I \subset [R, +\infty[$. The hypotheses of the mean value inequality are verified, namely:
\begin{itemize}
    \item[--] $f'$ is continuous over $[a, b]$;
    \item[--] $f'$ is differentiable over $]a, b[$;
    \item[--] for all $x \in ]a, b[$, $f''(x) \leq \epsilon \delta / 2$ (since $x \geq M$);
\end{itemize}
so that
$$ \left| \frac{f'(b) - f'(a)}{b-a} \right| \leq \frac{\delta}{2} \epsilon \, . $$
Yet, by definition of $I$, $f'(a) = \delta$, $f'(b) = \delta/2$ and $(b-a) = |I|$. The above inequality therefore boils down to
$$ \frac{\delta - \frac{\delta}{2}}{|I|} \leq \frac{\delta}{2} \epsilon \quad \iff \quad \frac{1}{|I|} \leq \epsilon \iff |I| \geq A \, ,$$
which concludes the proof.
\begin{enumerate}
\setcounter{enumi}{3}
    \item Contradiction.
\end{enumerate}
Finally, we use the convergence of $f$ to bring out a contradiction. Let $\epsilon > 0$ and $M \geq 0$ the constant associated to property (\ref{eqn:f_conv}). According to the previous point, there exists $I \in (I_n)_{n \in \mathbb{N}}$ such that
$$ \begin{cases}
I \subset [M, +\infty[ \\[5pt]
|I| \geq \dfrac{4}{\delta} \epsilon
\end{cases} \, . $$
Let us denote $[a, b] \coloneqq I$. On the one hand, the convergence of $f$ provides the inequality
$$ |f(b) - f(a)| \leq \epsilon \quad \text{because } a,b \geq M \ , $$
and on the other hand, $\forall x \in [a,b]$, $f'(x) \geq \delta/2$ so that the mean value inequality gives
$$ \int_a^b f'(x) \, \mathrm{d}x \geq \int_a^b \frac{\delta}{2} \, \mathrm{d}x \text{ hence } |f(b)-f(a)| \geq f(b) - f(a) \geq \frac{\delta}{2} |I| \geq \frac{\delta}{2} \frac{4}{\delta} \epsilon = 2 \epsilon > 0  \, .$$
The contradiction is clear. Q.E.D.

\section{Three-dimensional Laplacian operator}

\subsection{Spherical coordinates $(r, \theta, \varphi)$}

\begin{equation}
    \Delta_{\mathrm{Sp}}^{\mathrm{3D}} f = \frac{1}{r^2} \frac{\partial}{\partial r} \left( r^2 \frac{\partial f}{\partial r} \right) + \frac{1}{r^2 \sin (\theta)} \frac{\partial}{\partial \theta} \left( \sin (\theta) \frac{\partial f}{\partial \theta} \right) + \frac{1}{r^2 \sin^2(\theta)} \frac{\partial^2 f}{\partial \varphi^2}
\label{eqn:spherical_laplacian}
\end{equation}

\subsection{Cylindrical coordinates $(\rho, \varphi, z)$}

\begin{equation}
    \Delta_{\mathrm{Cyl}}^{\mathrm{3D}} f = \frac{1}{\rho} \frac{\partial}{\partial \rho} \left(\rho \frac{\partial f}{\partial \rho} \right) + \frac{1}{r^2} \frac{\partial^2 f}{\partial \varphi^2} + \frac{\partial^2 f}{\partial z^2}
\label{eqn:cylindrical_laplacian}
\end{equation}

\section{Line Search algorithm}
\label{sec:linesearch}

A line search algorithm can be implemented so as to compute an optimal value of the relaxation parameter at each iteration of the Newton method. To that extent, let us introduce the continuous functionals

\begin{flalign}
  \text{Let } v \in V, \ \begin{aligned}[t]
  f_v \colon V &\to \mathbb{R}\\
  u &\mapsto \alpha \int_{\Omega} \boldsymbol{\nabla}u \cdot \boldsymbol{\nabla}v \, \mathrm{d}x - \int_{\Omega} u^{-(n+1)} v \, \mathrm{d}x + \int_{\Omega} \rho v \, \mathrm{d}x \, .
\end{aligned} &&
\label{eqn:continuous-nonlinear}
\end{flalign}
\begin{flalign}
  \text{Let } v, \phi \in V, \ \begin{aligned}[t]
  \Tilde{f}_{v, \phi} \colon V &\to \mathbb{R}\\
  u &\mapsto \alpha \int_{\Omega} \boldsymbol{\nabla}u \cdot \boldsymbol{\nabla}v \, \mathrm{d}x +(n+1)\int_{\Omega} \phi^{-(n+2)} u v \, \mathrm{d}x - (n+2) \int_{\Omega} \phi^{-(n+1)} v \, \mathrm{d}x \int_{\Omega} \rho v \, \mathrm{d}x
\end{aligned} &&
\label{eqn:continuous-linear}
\end{flalign}
as well as their discrete counterpart
\begin{flalign}
  \begin{aligned}[t]
  F \colon \mathbb{R}^N &\to \mathbb{R}^N\\
  \mathbf{U} &\mapsto \Big(\alpha \sum_{j=1}^N U_j \int_{\Omega} \boldsymbol{\nabla}w_j \cdot \boldsymbol{\nabla}w_i \, \mathrm{d}x - \int_{\Omega} (u^h)^{-(n+1)} w_i \, \mathrm{d}x + \int_{\Omega} \rho w_i \, \mathrm{d}x \Big)_{1 \leq i \leq N}
\end{aligned} &&
\label{eqn:discrete-nonlinear}
\end{flalign}

\begin{flalign}
  \text{Let } \phi^h \in V^h, \ \begin{aligned}[t]
  \Tilde{F}_{\phi^h} \colon \mathbb{R}^N &\to \mathbb{R}^N\\
  \mathbf{U} &\mapsto \Big(\alpha \sum_{j=1}^N U_j \int_{\Omega} \boldsymbol{\nabla}w_j \cdot \boldsymbol{\nabla}w_i \, \mathrm{d}x + (n+1)\sum_{j=1}^N U_j \int_{\Omega} (\phi^h)^{-(n+2)} w_i w_j \, \mathrm{d}x \\ &- (n+2) \int_{\Omega} (\phi^h)^{-(n+1)} w_i \, \mathrm{d}x + \int_{\Omega} \rho w_i \, \mathrm{d}x \Big)_{1 \leq i \leq N}
\end{aligned} &&
\label{eqn:discrete-linear}
\end{flalign}
where, again, we recall that $u^h = \sum_{i=1}^N U_i w_i$. Now consider the $(k+1)^{\mathrm{th}}$ iteration of the Newton method. We know $\mathbf{U}^k = (U_1^k, \dots, U_2^k)^T$ from the previous iteration and we have computed $\mathbf{U}^*$ as the solution of the linear system $\Tilde{F}_{u^{h, k}}(\mathbf{U}) = \mathbf{0}$. It follows that the vector $\delta \mathbf{U} \coloneqq \mathbf{U}^* - \mathbf{U}^k$ is a direction of descent of $F_{u^{h, k}}$ so that the new iterate can be constructed as $\mathbf{U}^w = \mathbf{U}^k + w \delta \mathbf{U} = w \mathbf{U}^* + (1-w) \mathbf{U}^k$. Our goal is then to determine $w$ such that $\| F_{u^{h, w}} (\mathbf{U}^w) \|^2$ is minimal. Since
$$ \| F_{u^{h, w}} (\mathbf{U}^w) \|^2 = \sum_{i=1}^N \left[F_{u^{h, w}} (\mathbf{U}^w)\right]_i^2 \, , $$
we can examine each term $\left[F_{u^{h, w}} (\mathbf{U}^w)\right]_i$ separately. Moreover,
$$ u_{h, w} \coloneqq \sum_{j=1}^N U_j^w w_j = \sum_{j=1}^N (U_j^k + w \delta U_j) w_j \implies \frac{\mathrm{d}u_{h, w}}{\mathrm{d}w} = \sum_{j=1}^N \delta U_j w_j = \delta u_h \, . $$
In order to compute this minimum, one takes the derivative of $\| F_{u^{h, w}} (\mathbf{U}^w) \|^2$ with respect to $w$.
\begin{flalign*}
\frac{\mathrm{d}}{\mathrm{d}w}\left( \left[F_{u^{h, w}} (\mathbf{U}^w)\right]_i^2 \right) &= 2 \left[F_{u^{h, w}} (\mathbf{U}^w)\right]_i \times \Big[ \alpha \sum_{j=1}^N \delta U_j \int_{\Omega} \boldsymbol{\nabla} w_j \cdot \boldsymbol{\nabla} w_i \, \mathrm{d}x \\
 &+ (n+1) \int_{\Omega} \delta u_h \left( u_{h, w} \right)^{-(n+2)} w_i \, \mathrm{d}x \Big] \, .
\end{flalign*}
The full result is then simply twice the dot product between 
\begin{itemize}
    \item[--] $F_{u^{h, w}} (\mathbf{U}^w)$ and
    \item[--] $\mathbf{G}_w$ with $G_i^w \coloneqq \alpha \sum_{j=1}^N \delta U_j \int_{\Omega} \boldsymbol{\nabla} w_j \cdot \boldsymbol{\nabla} w_i \, \mathrm{d}x + (n+1) \int_{\Omega} \delta u_h \left( u_{h, w} \right)^{-(n+2)} w_i \, \mathrm{d}x $
\end{itemize}
Finally, we numerically find a zero of $:w \mapsto \mathbf{G}_w^T \ F_{u^{h, w}} (\mathbf{U}^w)$, e.g. via the regula falsi method.

\bibliographystyle{apsrev4-2}
\bibliography{references}

\end{document}